\documentclass[a4,10]{article}

\pdfoutput=1 

\usepackage{jheppub} 

\usepackage[T1]{fontenc} 
\makeatletter
\newenvironment{tablehere}
  {\def\@captype{table}}
  {}

\newenvironment{figurehere}
 {\def\@captype{figure}}
  {}
\makeatother

\usepackage{graphicx,times}
\usepackage{natbib}
\usepackage{amsmath}
\usepackage{booktabs}
\usepackage{diagbox}
\usepackage{indentfirst}
\usepackage{booktabs}
\usepackage{threeparttable}
\usepackage{array}
\usepackage{multicol}

\newcommand{\xmm}{{\it\,XMM-Newton }}
\newcommand{\nustar}{{\it\,NuSTAR }}
\def\ltsima{$\; \buildrel < \over \sim \;$}
\def\simlt{\lower.5ex\hbox{\ltsima}}
\def\gtsima{$\; \buildrel > \over \sim \;$}
\def\simgt{\lower.5ex\hbox{\gtsima}}
\bibpunct{(}{)}{;}{a}{}{,}

\title{\boldmath Response of the Fe K$\alpha$ line emission to the X-ray continuum variability in the changing-look active galactic nucleus NGC 1566}

\author[a]{W. C.~Liang,}
\author[a]{X. W.~Shu,}
\author[b]{J. X.~Wang}
\author[c]{ Y. Tan}
\author[a]{ W. J. Zhang}
\author[a]{L. M. Sun}
\author[b]{N. Jiang}
\author[d]{L. M.~Dou}

\affiliation[a]{Department of Physics, Anhui Normal University, Wuhu, Anhui, 241002, China}
\affiliation[b]{
              CAS Key Laboratory for Researches in Galaxies and Cosmology, Department of Astronomy, University of Science and Technology of China, Hefei, Anhui 230026, China}
\affiliation[c]{ Key Laboratory of Particle Astrophysics, Institute of High Energy Physics, Chinese Academy of Science, Beijing 100049, China}
\affiliation[d]{ Department of Astronomy, Guangzhou University, Guangzhou 510006, China}

\emailAdd{xwshu@mail.ahnu.edu.cn}

\abstract{  
NGC 1566 is a changing look AGN known to exhibit recurrent X-ray outbursts with each lasting for several years. 
The most recent X-ray outburst is observed on 2018, with a substantial increase of 2--10 keV flux by a factor of $\sim$24 than the historical minimum. 
We re-analyze the \xmm and \nustar observations covering the pre-outburst, outburst and post-outburst epochs, 
and confirm the discovery of the broad feature in the $\sim$5--7 keV band during the period of outburst that could be interpreted as a relativistic Fe K$\alpha$ emission line. 
Our analysis suggests that its flux has increased in tandem with the 2--10 keV continuum, making it the second changing look AGN 
in which the broad Fe K$\alpha$ line responds to the X-ray continuum variability. 
{This behavior strongly supports the idea that X-rays originates in a corona above the accretion disk, 
and disk reflection produces the relativistic Fe K$\alpha$ line. }
{In addition}, we find the response of narrow Fe K$\alpha$ emission line to the changes in the X-ray continuum on a time-scale as short 
as four months, allowing to put the location of line-emitting region at $<$0.1 pc, comparable to the size of optical BLR.  
By comparing to the changing look AGN NGC 2992, the Fe K$\alpha$ variation rate (the ratio of Fe K$\alpha$ variation to luminosity variation) in NGC 1566 appears greater, 
which could be possibly explained by larger amount of gas or Fe abundance responsible for producing the Fe K$\alpha$ line for the latter. 
{The strength of variable broad Fe K$\alpha$ line as well as the soft X-ray excess emission appears to be correlated with the accretion rate, 
which could be explained as due to the state transition associated with the changing-look phenomenon. }
\keywords{galaxies:active--galaxies:Seyfert--line:profiles--X-ray:individual (NGC 1566)}
}

\begin{document}
\maketitle
\flushbottom

\begin{multicols}{2}
\section{Introduction}
The relativistic Fe K$\alpha$ emission line in Active Galactic Nuclei (AGNs) is believed to originate from the innermost region of accretion disk 
by reflecting the illuminating X-ray continuum, for which the line profile is significantly broadened and skewed towards lower energies due to 
the relativistic Doppler effect and strong gravitational redshift \citep[e.g.,][]{1989MNRAS.238..729F, 1991ApJ...376...90L}. 
Therefore, proper modeling the Fe K$\alpha$ line profile allows for constraining the inclination angle of the disk, the emissivity of the disk and its ionization state, 
as well as the spin of the supermassive black hole (SMBH) \citep[e.g., ][]{Nandra2007, delacalle2010, Patrick2012, Reynolds2014, 2012ApJ...747L..11T}. 
While there are plenty of studies investigating the broad Fe K$\alpha$ component by modeling of the time-averaged spectrum in literature 
\citep[see the reviews by][]{Fabian2000, Miller2007, Reynolds2021}, 
{spectral studies on the long-term variability of the broad Fe K$\alpha$ line on timescale of years are relatively limited 
\citep[e.g.,][]{2004MNRAS.355.1073I, 2007A&A...470..889P, 2009A&A...507..159D, 2016ApJ...832...45N, 2020MNRAS.496.3412M}. 
In principle, the variability of the line profile, especially its relation to the variability of the continuum,
can provide critical information on the accretion disk and black hole parameters in more model-independent ways \citep{2006A&A...445...59T, 2007A&A...467.1057T, 2018MNRAS.478.5638M}. 
On shorter timescale of minutes, X-ray reverberation mapping has been proven possible and provided a dynamic view of the inner accretion region 
\citep[e.g.,][]{Zoghbi2010, Parker2014, Alston2020, Wilkins2021}. 
\citet{2016MNRAS.462..511K} presented a statistical study of X-ray time lags in a sample of 43 Seyfert galaxies, and found that approximately 50 per cent of 
sources exhibit Fe K reverberation in which the X-ray emission in Fe K band (6-7 keV) responds to rapid variability in the continuum. 
While the Fe K reverberation is powerful to spatially map the geometry of the inner accretion flow, only the time-lag information of 
emission between different bands can be obtained. 

On longer timescales of days to years, very few sources have been reported in which the variability behavior of the broad Fe K line relative to the X-ray continuum has been investigated, 
e.g., MCG-6-30-15 \citep{Ballantyne2002, Marinucci2014}, NGC 4051 \citep{Wang1999, 2006MNRAS.368..903P}, 1H 0707-495 \citep{Fabian2004, Fabian2012}, 
IRAS 13224-3809 \citep{Fabian2013, Jiang2018} and NGC 2992 \citep{Murphy2007}. Among them, NGC 2992 is the only source which shows the unambiguous 
evidence that the broad Fe K line flux is responding to the long-term continuum variation over decades \citep{2010ApJ...713.1256S, 2018MNRAS.478.5638M, 2020MNRAS.496.3412M}. 
}
{In some AGNs, the disk reflection component and the broad Fe K$\alpha$ line (if present) 
are found to vary with much smaller amplitude than the continuum \citep{Markowitz2003}, which can be interpreted in the context of the light-bending model. 
In such a model, }physical motion of the X-ray source causes variability in the X-ray continuum but the illumination of the disk (and
therefore the Fe K$\alpha$ line flux) {can remain relatively unaffected in some regimes \citep{Fabian2004, Miniutti2004}. }
The observation of broad Fe K$\alpha$ variability in response to {the large-amplitude continuum variability in NGC 2992 suggests that 
the light-bending scenario may not be relevant} for at least this AGN. 
The mechanisms that cause large-amplitude X-ray continuum variability in AGNs remains further explored. 


The time-domain spectral observations have revealed a class of AGNs which can change their types on time scales of years, 
characterized by transient emergence or disappearance of broad emission lines 
\citep[e.g.,][]{Shappee2014, LaMassa2015, MacLeod2016}.
These objects are so-called changing-look (CL) AGNs. Follow-up studies of these systems suggested that a
dramatic change in the accretion rate, rather than change in the obscuration, is the most likely explanation
of their spectral transitions \citep{Runnoe2016, Sheng2017}. 
Large amplitude X-ray variability has also been observed to correlate with the CL phenomenon \citep{Denney2014, Husemann2016, Mathur2018, Guolo2021}.  
In this case, the accretion disk can be subject to large variability in illumination by the
X-ray continuum and therefore produces an variable Fe K$\alpha$ emission, providing 
us with a unique testbed to study the response of the broad Fe K$\alpha$ emission line to the changes 
in the X-ray continuum. 

 \vspace{0.6cm}
\begin{figurehere}
\centering
\includegraphics[scale=0.25]{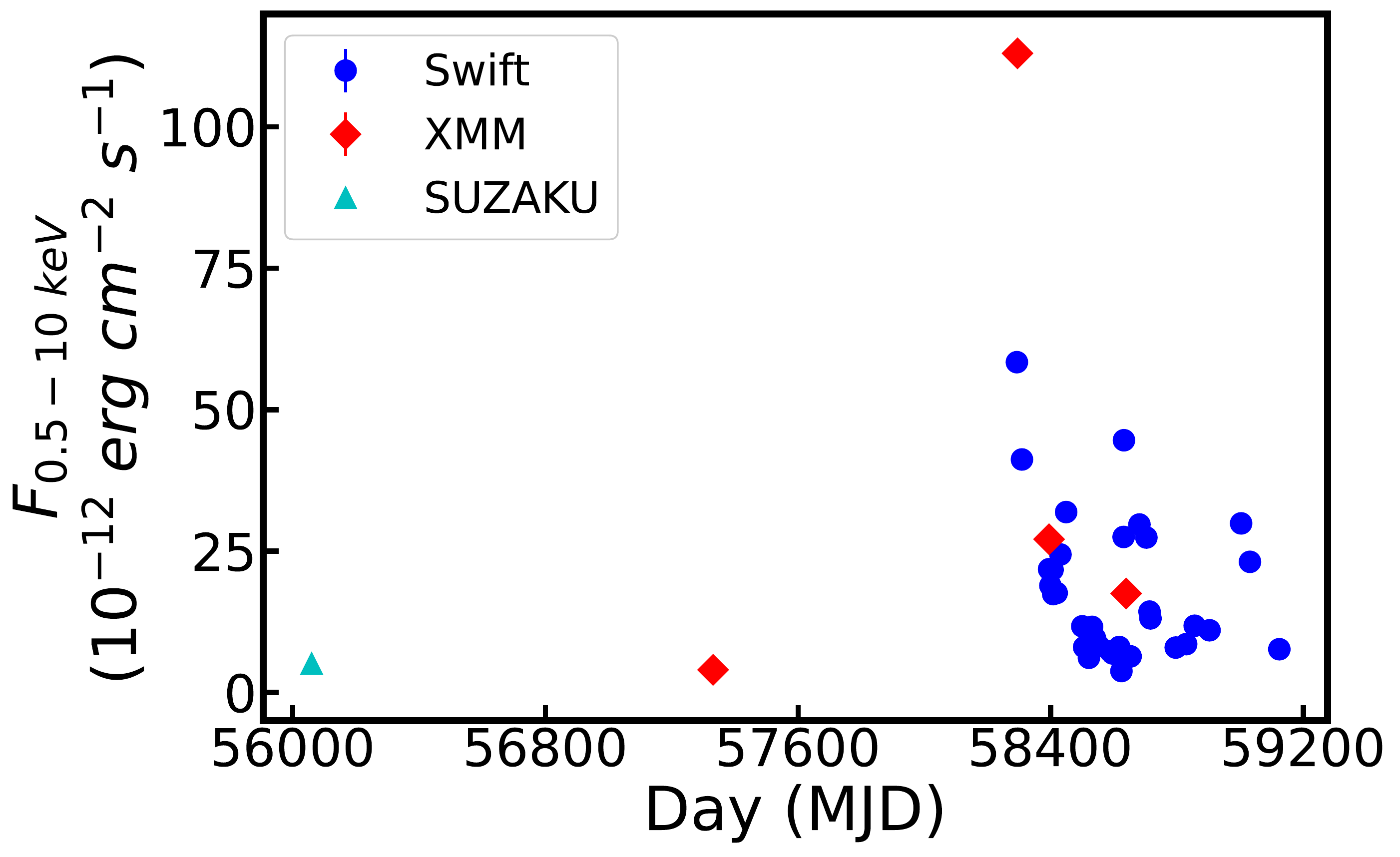}
\caption{\label{Fig:lcurve} The historical 0.5-10.0 keV light curve of NGC 1566 since 2012. 
The data for \textit{Suzaku} are taken from \citet{Kawamuro2013}. The \xmm and {\it Swift}/XRT data are reduced by 
our work, followed the procedures given in \citet{Shu2018, Shu2020}. }
\end{figurehere}
 \vspace{0.6cm}

NGC 1566 is a nearby AGN at a redshift $z=$0.00502 \citep{2004AJ....128...16K}. 
\citet{2019MNRAS.483..558O} classified it as a new CL AGN based on the type changes in optical spectra and flux variations in the past few decades. 
In the X-ray band, the studies before 2013 mainly focused on the spectral properties below 6 keV \citep{1990ApJ...361..459E,1992A&A...256..375B}.
 \citet{Kawamuro2013} performed the first broadband X-ray spectral analysis based on the data taken with \textit{Suzaku} and \textit{Swift/BAT}. 
 There is a strong narrow Fe K$\alpha$ line component (EW $=$ 240 eV) in the broadband spectra, but no obvious broad Fe K line (EW $<$ 32 eV) emission is detected. 
 Since its discovery as a CL AGN, \citet{2019MNRAS.483..558O} reported the huge UV/optical outbursts from the coincident \textit{Swift} observations. 
 More importantly, the X-ray continuum flux increased by nearly 50 times from the lowest state in 2014 to the highest state recorded in July 2018.  
 The quasi-simultaneous \textit{XMM-Newton} and \textit{NuSTAR} observations have revealed complex components in the X-ray spectrum \citep{2019MNRAS.483L..88P}, 
 including a strong and variable soft excess, warm absorption, Compton hump and reflection components, and Fe K$\alpha$ emission. 
 \citet{Jana2021} suggested that the increase in the accretion rate is responsible for the sudden rise in luminosity, and NGC 1566 
 might harbour a pair of merging supermassive black holes. 
 Although the broadband spectral properties have been investigated in previous works \citep{2019MNRAS.483L..88P, Jana2021}, 
 to the best of our knowledge no study has yet explored the link of the variability of Fe K$\alpha$ line with the X-ray continuum, 
 nor attempted to investigate in detail the behaviour of Fe K$\alpha$ line variability by decomposing it into broad and narrow components. 
 These are our goals in this paper. 
 According to the criteria proposed by \citet{2010ApJ...713.1256S}, NGC 1566 may represent one of few AGNs {with large accretion rate change}, in which the Fe K$\alpha$ 
 line emission in response to the large amplitude continuum variability can be studied, thanks to its identification as a new CL AGN with intensive multi-wavelength outbursts. 
 The paper is organized as follows. In Section 2 we describe the observations and data reduction. In Section 3, we present the X-ray spectral analysis and the detailed results of the Fe K$\alpha$ line variability by comparing the data between different X-ray states. In Section 4, we discuss some general implications of our findings and present our conclusions.
\section{Observations and Data Reduction}

In order to investigate the broad Fe K$\alpha$ line and its variability behaviors, sensitive X-ray observations are required to 
obtain time-sliced spectra with sufficient signal-to-noise ratio. This is crucial to detect the broad Fe K$\alpha$ line and 
constrain the line profile parameters with sufficient accuracy. Although NGC 1566 has been observed intensively by {\it Swift}/XRT, 
in this paper we focus mainly on the data taken from the \xmm observations, and quasi-simultaneous \nustar observations if available. 
Details of the \xmm and \nustar observations are presented in Table 1. 
As shown in Figure 1, there are four \xmm observations, which cover the periods of X-ray low state, outburst state and decline 
state\footnote{There are totally five archival \xmm observations, but only four have sufficient spectral quality to perform 
the analysis of Fe K$\alpha$ line emission.}
. Therefore, the observations are suitable for our study. 

 \vspace{0.5cm}

\begin{tablehere}
\centering
\setlength{\tabcolsep}{0.095mm}{
\begin{tabular}{ccccc}\hline\hline
Mission$^a$ & Obs No. & ObsID & Obs Date$^b$ & 	Exposure \\
\hline
XMM & Obs-1 & 0763500201  & 2015-11-05 & 91.9ks \\
XMM & Obs-2 & 0800840201  & 2018-06-26 & 94.2ks \\
XMM & Obs-3 & 0820530401  & 2018-10-04 & 108ks \\
XMM & Obs-4 & 0840800401  & 2019-06-05 & 94.0ks \\
Nu  & Obs-1 & 80301601002 & 2018-06-26 & 56.84ks \\
Nu  & Obs-2 & 80401601002 & 2018-10-04 & 75.4ks \\
\hline
\end{tabular} }
\caption{Observation details of NGC 1566. $^a$XMM:\textit{XMM-Newton}; Nu:\textit{NuSTAR}. $^b$The dates are reported in the year-month-day format.\label{Tab:Obs-list}}
\end{tablehere}
 \vspace{0.5cm}

\begin{figurehere}
\centering
\includegraphics[scale=0.1]{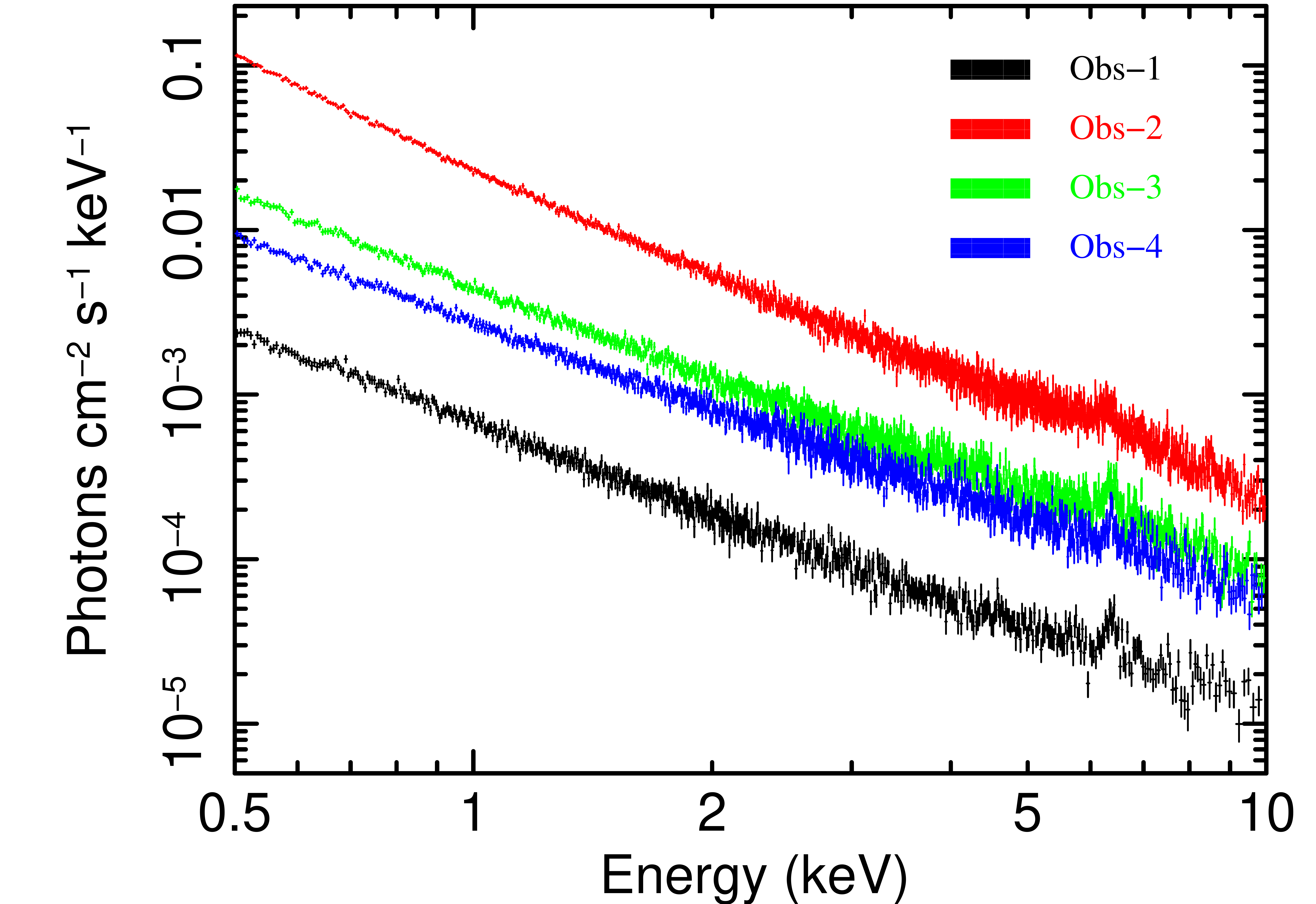}
\caption{\label{Fig:allspec} 
{Comparison between the X-ray spectrum of NGC 1566 observed with \xmm before, 
during and after the outburst (Table 1). }
}
\end{figurehere}

\begin{figure*}[htbp!]
\centering
\includegraphics[scale=0.21]{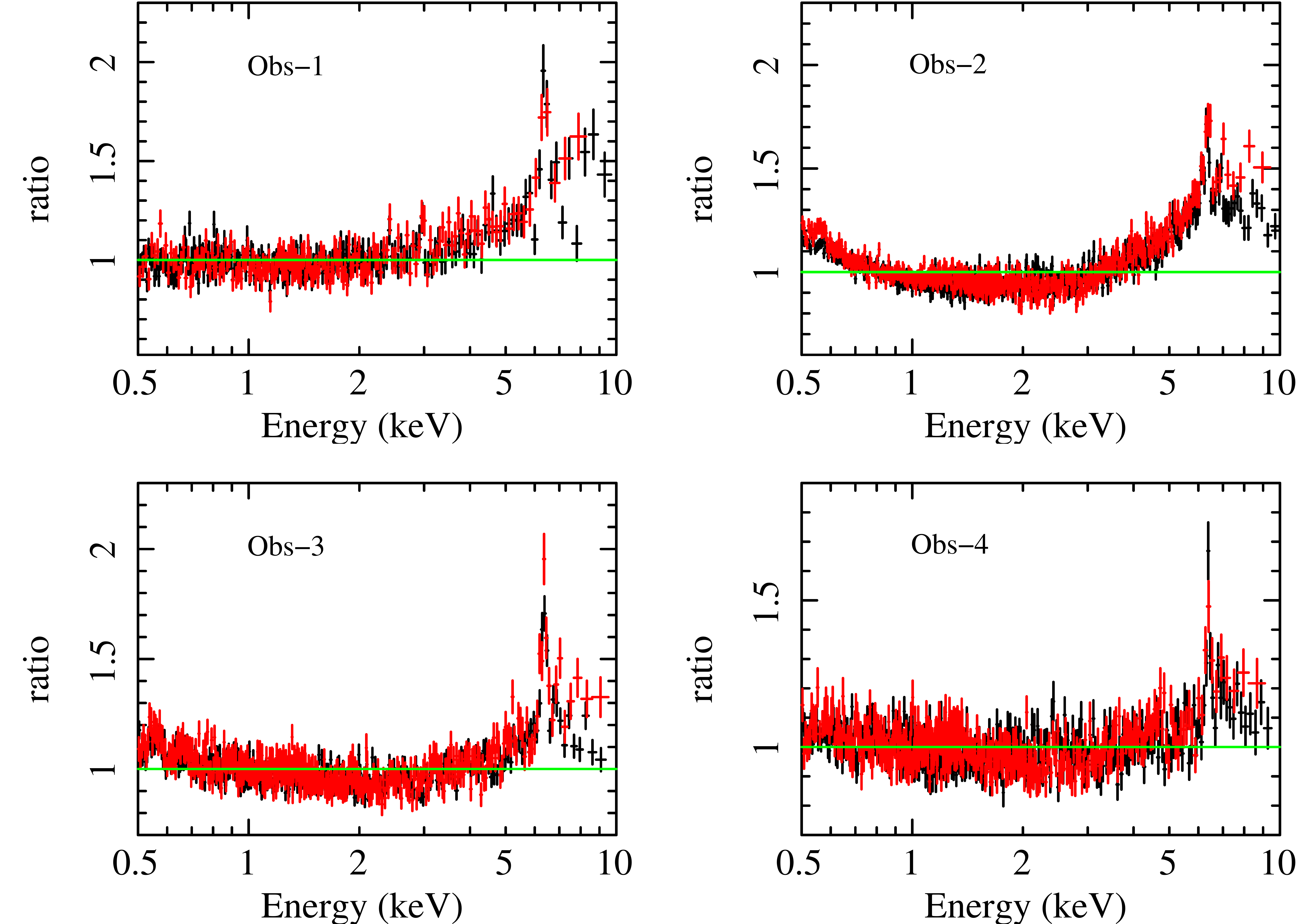}
\caption{\label{Fig:xmm_p_m_ra} Ratios of 0.5--10.0 keV spectra from four \xmm observations to a model consisting of a simple, absorbed power-law continuum (see Section \ref{Sec:pre_fit}). 
Black and red data points correspond to the pn and summed MOS1 and MOS2 data, respectively.}
\end{figure*}

\subsection{\xmm}
We reduced the data from \textit{XMM-Newton} observations following the standard procedures using the \textit{XMM-Newton} Science Analysis System (SAS) v18.0.0. 
For the raw event files of EPIC pn and MOS detectors, we produced clean event files by using {\tt EPCHAIN} and {\tt EMCHAIN} respectively. 
Science products were then produced using XMMSELECT. The source spectra were extracted from a circular region with $40^{\prime\prime}$ radius centered on the source position and the extent of pile-up effect was assessed by using the SAS task \texttt{epateplot}. We found that the last three observations from both \textit{XMM-Newton} EPIC pn and MOS data at the core of the point-spread function for NGC 1566 were affected by photon pile-up.

We then used \texttt{epatplot} to compare the spectra of the last three observations that were extracted in annular regions with different 
inner radius, while the outer radius was fixed at $40^{\prime\prime}$. Therefore, by effectively mitigating the pile-up effect, 
the extraction region of the source spectrum is: circle ($40^{\prime\prime}$),
annuli ($15^{\prime\prime}$-$40^{\prime\prime}$), annuli ($10^{\prime\prime}$-$40^{\prime\prime}$), annuli ($7.5^{\prime\prime}$-$40^{\prime\prime}$) 
for Obs-1, Obs-2, Obs-3, Obs-4, respectively. The background was extracted from two circular regions of blank sky on the same detector for pn, while it was from 
four circular regions for MOS detector, with each circle having a radius of $40^{\prime\prime}$. 
Instrumental response files were generated with {\tt RMFGEN} and {\tt ARFGEN} for each detector. After performing the 
spectral extraction separately for MOS1 and MOS2, we combined the data into a single spectrum using {\tt ADDASCASPEC}. 
{Figure \ref{Fig:allspec} shows the comparison of pn spectrum observed at different epochs. }

\subsection{\nustar}
The \textit{NuSTAR} data were processed using v.1.6.0 of the
{\tt NuSTARDAS} pipeline with CALDB v20160303. For each observation, we extracted source counts from a circular region with $40^{\prime\prime}$ radius 
centered on the source position, and background counts from a source-free circular region with a radius of $50^{\prime\prime}$ on the same chip. 
The spectra were generated using the {\tt NUPRODUCTS} task {and grouped to have at least 40 counts in each bin for following spectral analysis}.


\section{Spectral Analysis}

\subsection{Preliminary Spectral Fitting}\label{Sec:pre_fit}
In order to characterize the main spectral components of the X-ray spectra, we first fitted the four \textit{XMM-Newton} pn spectra 
between 0.5-10.0 keV using {the spectral fitting software {\tt XSPEC} \citep[Version 12.11.1,][]{1996ASPC..101...17A}, by adopting} a simple absorbed power-law model with the photon index $\Gamma$ allowed to float. 
{We grouped the spectra to have at least 40 counts in each bin so as to adopt the $\chi^2$ statistic for the spectral fits}. 
The Galactic column density  
at $N_H = 7.11 \times 10^{19}\, cm^{-2}$ is fixed during all the model fitting processes. All statistical errors of spectral model parameters are 
reported at the confidence 90$\%$ for one parameter ($\Delta\chi^2=2.706$). We will give spectral parameters and statistical errors below, 
when the complete fitting results to the data are obtained. As shown in Figure \ref{Fig:xmm_p_m_ra}, the ratios of the data to model show that 
the spectra display a variable soft X-ray excess below 2 keV which strongest in strength during the second \xmm observation (2018 Jun 26). 
Meanwhile there is also a considerable curvature in the spectrum between $\sim$ 4-8 keV. We know from previous observations of NGC 1566 that 
the residuals in the range of $\sim$ 4-8 keV can be interpreted as the prominent Fe K line emission\citep{Kawamuro2013, 2019MNRAS.483L..88P}. 
Figure \ref{Fig:xmm_p_m_ra} also shows the ratio of the summed MOS1 and MOS2 data to the same model. 
It can be seen that the MOS data are consistent with the pn data within the statistical errors, 
{though there is a small discrepancy between the MOS and pn at high energies in Obs-2 and Obs-3. 
This is possibly due to the imperfect correction for the effects of pile-up, as the effects are worse for the MOS than the pn. 
Since we used principally the pn data, which have much higher sensitivity, such effects on the MOS data would  
not affect the following spectral analysis.}


\subsection{Baseline Model}\label{Sec:base_model}

\begin{figure*}[htbp!]
\centering
\includegraphics[scale=0.3]{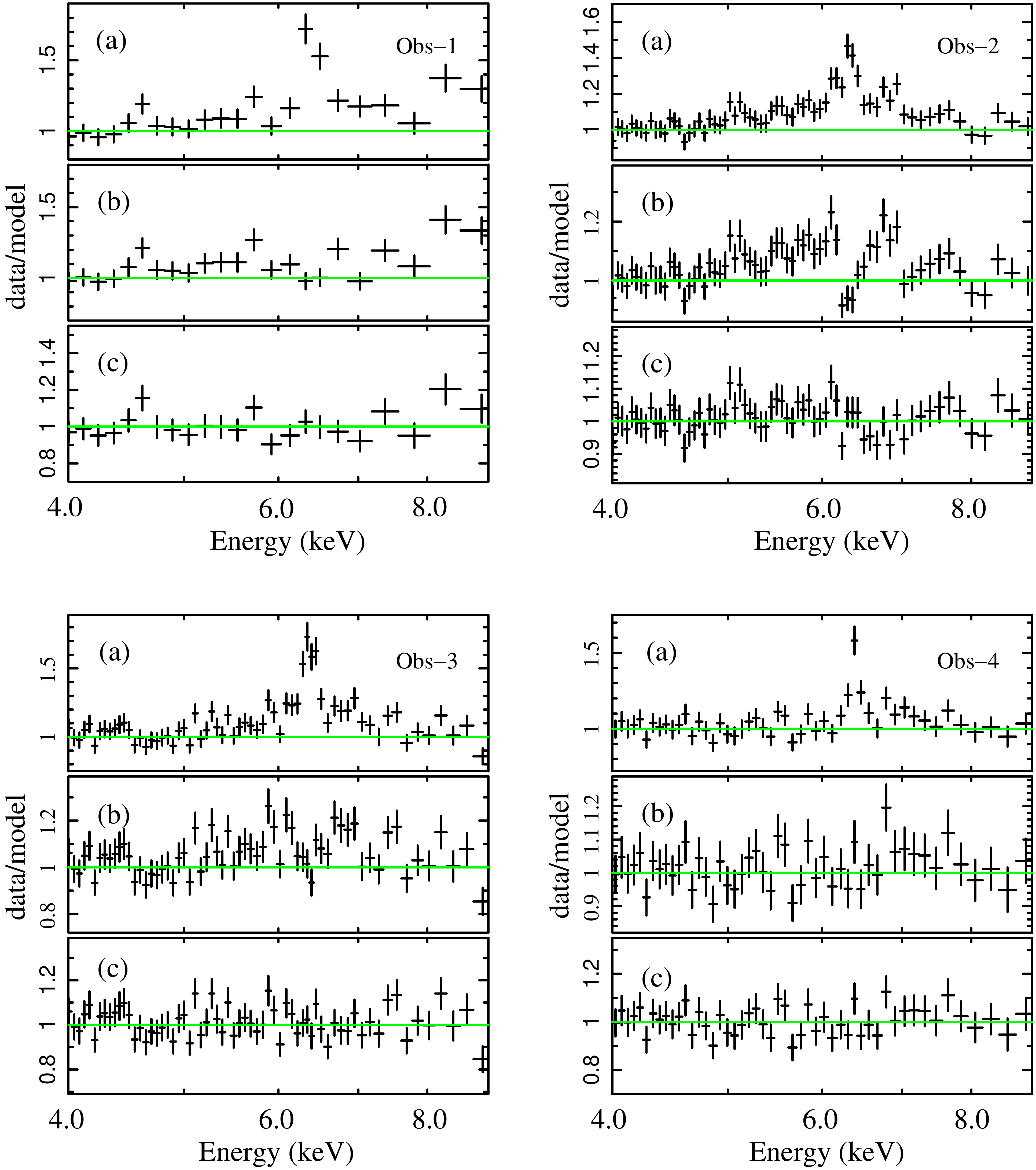}
\caption{\label{Fig:xmm_4-9ra} Data/model ratios for the NGC 1566\textit{ XMM-Newton} pn data in the 4-9 keV band. 
(a) Data/model ratio after fitting an dual power-law continuum (see Section \ref{Sec:pre_fit} and \ref{Sec:base_model} for details). 
(b) Residuals after the narrow Fe K$\alpha$  and Fe K$\beta$ lines have been fitted. An excess between 5-7 keV is still present. 
(c) Residuals after the addition of a broad Fe K$\alpha$ and K$\beta$ line, using the baseline model described in Section \ref{Sec:base_model} 
(see also Table \ref{Tab:xmm_result}).}
\end{figure*}

\begin{figure*}[htbp!]
\includegraphics[scale=0.3]{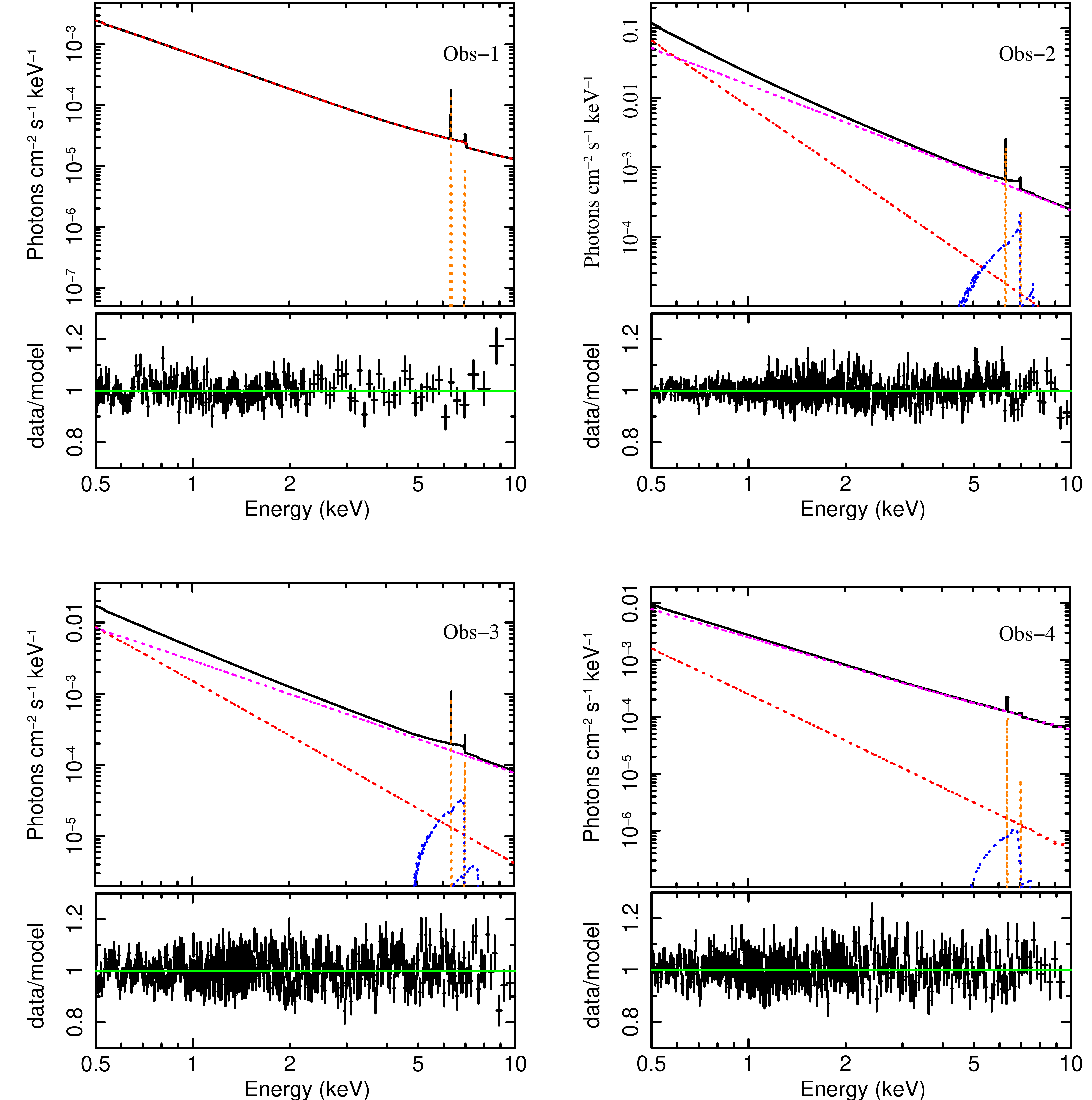}
\caption{\label{Fig:xmm_ufra} The best-fitting baseline model to the NGC 2992 \textit{XMM-Newton} data (top panel), and the corresponding data/model ratio (bottom panel). \textit{Red dotted line}: power-law model, used to explain soft X-ray excess components; \textit{Purple dotted line}: The pexrav model is used to explain the reflection continuum component; \textit{Blue dotted line}: The disk-line model is used to explain the broad Fe K line component; \textit{Brown dotted line}: The zgauss model is used to explain the narrow Fe K line components.}
\end{figure*}

{Next, we performed more detailed spectral fittings, in order to investigate the Fe K line complex in detail. }
As shown in Figure \ref{Fig:xmm_p_m_ra}(a), the data to model ratio from a simple absorbed powerlaw model suggests an excess emission towards 
higher energies, which is likely due to the reflection component \citep[e.g.,][]{Fabian2010}. We added a standard reflection model (\texttt{pexrav} in XSPEC, \citep{1996ASPC..101...17A}) 
to account for the Compton reflection emission in the \textit{XMM-Newton} data \citep{1995MNRAS.273..837M}. 
Although the effective energy band of \textit{XMM-Newton} (0.2-10.0 keV) does not constrain the reflection component well, it may have effect on the modelling of the Fe K emission lines. 
\texttt{pexrav} model calculates the reflected spectrum produced by an X-ray source illuminated onto an optically-thick neutral disk. 
During the spectral fittings, all model parameters were fixed to their canonical values except the reflection fraction ($R$), 
which defines the strength of the reflection component relative to that expected from a slab subtending 2$\pi$ solid angle. 
The reflection continuum in our baseline model is only an empirical parameter, as the \texttt{pexrav} model ignores fluorescence 
and the geometry is limited to a planar disk. In Section \ref{Sec:Var_ref}, we will present a more self-consistent modeling of the refection component 
as well as the Fe K$\alpha$ emission lines. 

In addition to the reflection model, we added another power-law component to account for the variable soft X-ray excess emission 
for the last three \xmm observations. 
Figure \ref{Fig:xmm_4-9ra} (a) shows the ratio of data to model between 4-9 keV. If the data in the 5.0-7.5 keV band were ignored, where the Fe K line profile would dominate, 
 the fitting results are acceptable for the four observations with $\chi^2/dof=219.33/189$, $404.31/403$, $401.06/363$, and $324.08/323$, respectively. However, the fittings 
 become poorer when the data in the 5.0-7.5 keV band were noticed ($\chi^2/dof=366.51/201$, $651.31/436$, $648.46/399$, $407.65/350$). This indicates 
 the presence of excess emission the Fe K band. Then we added two Gaussian components to account for narrow Fe K$\alpha$ and Fe K$\beta$ line emission 
 from the distant matter, which is ubiquitous in the X-ray spectra of AGNs \citep[e.g.,][]{Shu2010b, Shu2011}. We tied the intrinsic line width $(\sigma_N)$ of Fe K$\beta$ for Fe K$\alpha$ and fixed it at 5 eV, which is unresolved with \textit{XMM-Newton} ($\sim$150 eV at 6 keV). The centroid energy $(E_N)$ and the intensity $(I_N )$ of Fe K$\alpha$  were allowed to vary. The Fe K$\beta$ line energy was fixed at 7.058 keV with the line flux equal to 13.5$\%$ of the Fe K$\alpha$ as expected for neutral Fe \citep{2004ApJS..155..675K}. 
 The overall fit results of the four observations have been significantly improved ($\Delta\chi^2=$101.5, 117.81, 158.34 and 54, respectively) for 
 two extra free parameters.\par

\begin{table*}[htbp!]
\centering
\caption{\label{Tab:xmm_result} Spectral Fitting Results for NGC 1566 \textit{XMM-Newton} Data}
\begin{tabular}{ccccc}\hline\hline
Observation Date                    & 2015-11-05               & 2018-06-26                       & 2018-10-04               & 2019-06-05\\
Parameters                          &                          & value                            &                          & \\\hline
$\chi^2$ /d.o.f                     & 166/197                  & 450/430                          & 448/394                  & 350.4/345\\
    power-law, $\Gamma$ &......                                 & $3.21_{-0.12}^{+0.42}$           & $2.57_{-0.28}^{+0.38}$   & $2.72_{-0.50}^{+0.56}$\\
    pexrav, $\Gamma$                 & $1.92_{-0.016}^{+0.016}$ & $1.81_{-0.045}^{+0.13}$          & $1.58_{-0.14}^{+0.08}$   & $1.70^f$\\
    Reflection fraction, R          & $3.18_{-0.57}^{+0.60}$   & $<1.11$                          & $<1.24$                  & $<1.52$\\
    $E_N \rm [Fe~K\alpha] (keV)$       & $6.39_{-0.029}^{+0.030}$ & $6.34_{-0.028}^{+0.028}$         & $6.39_{-0.018}^{+0.021}$ & $6.41_{-0.026}^{+0.025}$\\
    Inner radius of disk, $R_{in}$  & $16^f$                   & $10.93(<16.5)$                   & $16.13(<29.03)$          & 6.86$(<38)$\\
    $\theta_{obs}$ (degrees)        & $42^f$                   & $41.37_{-1.91}^{+2.82}$          & $42.50_{-4.33}^{+10.88}$ & $42^f$\\
    $I_N \rm [Fe~K\alpha] (10^{-5}\, photons\, cm^{-2}\, s^{-1}$)  & $0.39_{-0.10}^{+0.10}$ & $3.64_{-0.93}^{+0.92}$ & $2.14_{-0.40}^{+0.40}$ &$1.10_{-0.29}^{+0.28}$\\
     $I_{disk} \rm [Fe~K\alpha] (10^{-5}\, photons\, cm^{-2} \,s^{-1})$ &$<0.11$            & $14.03_{-4.31}^{+2.33}$ & $3.70_{-0.96}^{+0.98}$ &$0.99_{-0.80}^{+0.82}$\\
     $EW_N \rm [Fe~K\alpha] (eV)$     & $139_{-35}^{+35}$        & $54_{-14}^{+14}$                 & $107_{-20}^{+20}$        & $86_{-22}^{+21}$\\
     $EW_{disk} \rm [Fe~K\alpha] (eV)$ & $<44$                    & $280_{-86}^{+44}$                & $230_{-60}^{+61}$        & $93_{-75}^{+75}$\\
     $F_{0.5-2.0keV}(10^{-11}\,erg \,cm^{-2}\, s^{-1})^a$      & 0.15                  & 5.31     & 1.00                     & 0.61\\
     $F_{2.0-10.0keV}(10^{-11}\, erg\, cm^{-2} \,s^{-1})^a$    & 0.25                  & 5.99     & 1.71                     & 1.14\\
     $L_{0.5-2.0keV}(10^{42} \,erg\, s^{-1})^b$                & 0.084                 & 2.96     & 0.56                     & 0.34\\
     $L_{2.0-10.0keV}(10^{42} \,erg\, s^{-1})^b$               & 0.138                 & 3.33     & 0.95                     & 0.64\\\hline
\end{tabular}
\begin{tablenotes}
\item[1]{\textbf{Notes}: Statistical errors and upper limits correspond to 90$\%$ confidence for one interesting parameter
($ \Delta\chi^2 $= 2.706). $^f$ The parameter was frozen to its most probable value as it can not be constrained in the spectral fittings.
 $^a$Observed-frame fluxes, \textit{not} corrected for Galactic and intrinsic absorption.
 $^b$Intrinsic, rest-frame luminosities, corrected for all absorption components.}
\end{tablenotes}
\end{table*}

Figure \ref{Fig:xmm_4-9ra} (b) shows that there is still some residual emission in the Fe K band, even if we have added two Gaussian lines 
to fit the narrow Fe K$\alpha$ and Fe K$\beta$ lines, revealing that the spectral fittings should take into account the broader component. 
The broader component could be the relativistic Fe K line originating from the line radiation in the inner area of the accretion disk. 
Alternatively, other factors can also lead to similar excess emission, such as complex absorption with different ionization states \citep[e.g.,][]{Miller2009, Patrick2012}. 
Considering the data quality of \textit{XMM-Newton} and the fact that NGC 1566 is a bare AGN with little absorption along the line of sight \citep{2002MNRAS.331..154R, Kawamuro2013}, the excess emission is {unlikely to be} caused by the complex absorption, and likely related to the relativistic Fe K line component. 

Therefore, we added a disk line ({\tt diskline} model in XSPEC) to account for the broad Fe K$\alpha$ emission line from the relativistic accretion disk around a Schwarzschild black hole \citep[see][]{1989MNRAS.238..729F}. The model consists of six free parameters: 
the line energy $E_0$, the power-law index of the line emissivity $q$, the inner and outer radius ($R_{in}$ and $R_{out}$, in units of $R_g=GM/c^2$), the inclination angle between the normal direction of the disc and the observer ($\theta_{obs}$), and the integrated intensity of the line ($I_{disk}$). 
As did in \citet{2010ApJ...713.1256S}, we fixed the $E_0$  at 6.4 keV in the rest frame, the emissivity at $q = -3$, and the outer radius $R_{out}$ at 1000$R_g$. 
Considering that the broad Fe K$\alpha$ line originates from the inner region of the accretion disk, we set the inner radius $R_{in}$ as a free parameter. 
The inclination ($\theta_{obs}$) and line intensity ($I_{disk}$) were also allowed to vary in the fittings. 
For consistence, we included an Fe K${\beta}$ component without any additional free parameters, and assume the same ratio as the Fe $K\beta/K\alpha$ of the narrow Fe K line. 
If the ionization degree of Fe is less than Fe$_{\rm XVII}$, Fe K$\beta$ will be produced \citep[e.g.,][]{2010ApJ...713.1256S}. 
Note that we fixed the inclination and/or inner radius of the Obs-1 and the Obs-4 at the mean value from the best-fit parameters of Obs-2 and Obs-3, 
because the broad Fe K line emission becomes relatively weak in the lower states, hence the above parameters cannot be well constrained. 
In summary, our baseline model can be written as:
\begin{flushleft}
{\small\textbf{$\bullet$constant*wabs*(pexrav+zgauss+zgauss+diskline+disk\\\quad -line) (for Obs-1)}}\\	
{\small\textbf{$\bullet$constant*wabs*(pexrav+zpowerlaw+zgauss+zgauss+d\\\quad -iskline+diskline) (for Obs-2, 3 and 4)}}\\
\end{flushleft}
The best-fitting parameters and their statistical errors are shown in Table \ref{Tab:xmm_result}, and will be discussed in detail 
in the following Section. 
Note that for {\it XMM} Obs-4, the photon index for the hard X-ray powerlaw component becomes extremely flat ($\Gamma=1.29^{+0.28}_{-0.62}$) 
when the extra powerlaw component is added to account for the soft-excess emission. Such a flat photon index appears not physical as it is not 
confirmed by the \nustar data (Section 3.4), and likely caused by the model degeneracy between the two powerlaw components. 
Therefore, in our baseline model for Obs-4, we fixed the hard X-ray powerlaw component at $\Gamma=1.7$, the best-fit value from the \nustar observations  
(Table \ref{Tab:rel_xill_result}). 
Figure \ref{Fig:xmm_ufra} shows the best-fitting baseline models for the four \xmm observations, along with the data to model residuals. 
{The data/model residuals for the same model are also shown in Figure \ref{Fig:xmm_4-9ra} (c) for just the 4-9 keV band in order to display the Fe K region more clearly.} 
\subsection{Broad versus Narrow Fe K Line}

\begin{figurehere}
\centering
\includegraphics[scale=0.16]{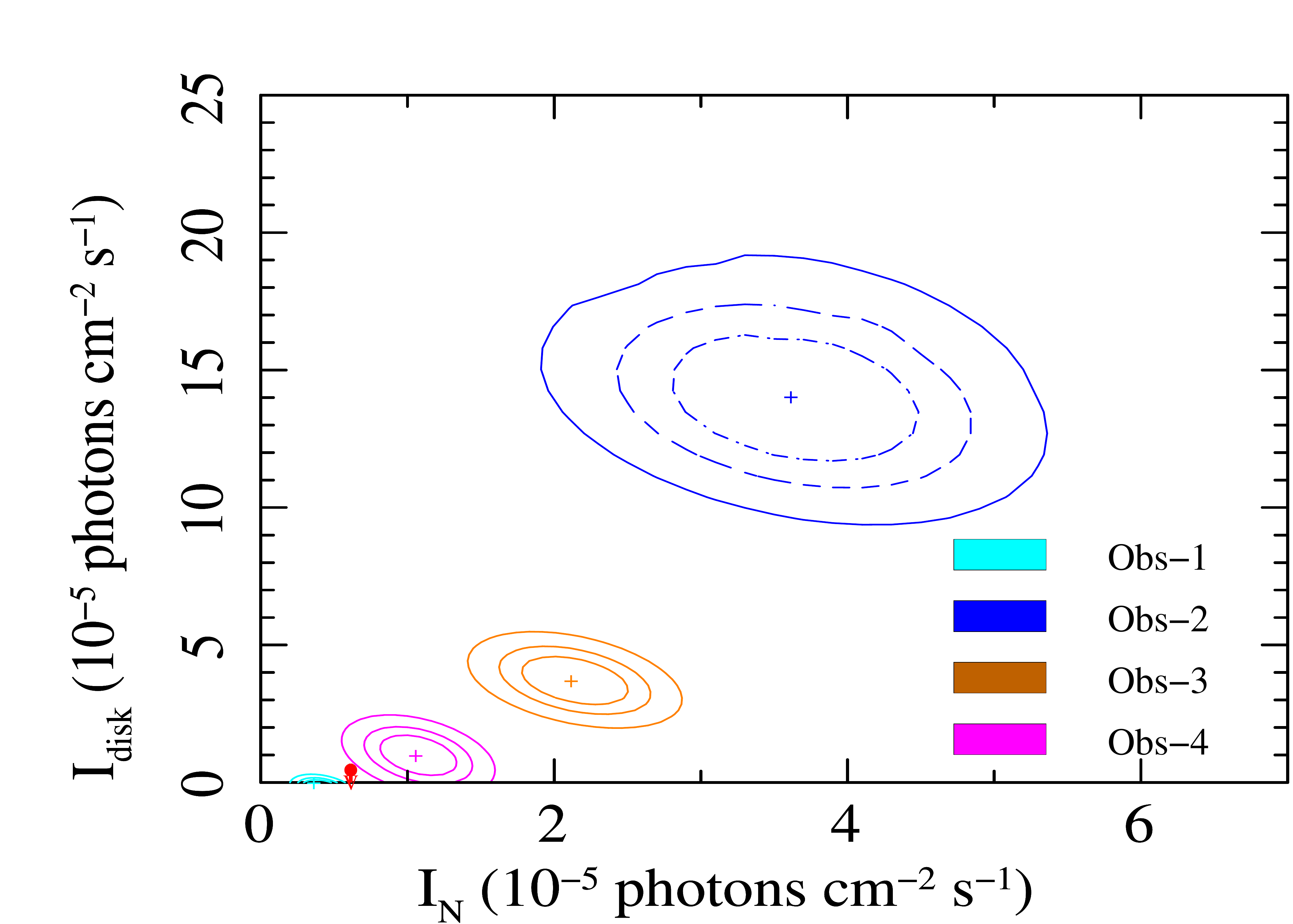}
\caption{\label{Fig:In_vs_Id_ct} Contour plot of the intensity of broad Fe K$\alpha$ line ($I_{\rm disk}$) versus narrow Fe K$\alpha$ line intensity ($I_{\rm N}$) for the NGC 1566 \textit{XMM-Newton} data. The contour from inside to outside represents 68$\%$, 90$\%$, and 99$\%$ confidence interval, respectively. The red dot indicates the broad and narrow Fe K$\alpha$ line intensity measured from the low-state {\it Suzaku} data \citep{Kawamuro2013}. }
\end{figurehere}

Following the work of \citet{2010ApJ...713.1256S}, we also investigated the extent to which the broad Fe K disk-line component and the narrow, 
distant-matter Fe K line may be decoupled based on the best-fitting models (see Figure \ref{Fig:xmm_ufra}) for the four \textit{XMM-Newton} observations. 
Figure \ref{Fig:In_vs_Id_ct} shows the contour plots between the narrow-line intensity ($I_N$ ) and the disk-line intensity ($I_ {disk}$). 
The line from inside to outside represents 68$\%$, 90$\%$, and 99$\%$ confidence level, respectively. 
It can see that the narrow Fe K$\alpha$ line intensity in all the \xmm observations 
is required to be non-zero at a high level of significance. Notably, in the context of the baseline model 
fitted here, the narrow K$\alpha$ line intensity in the highest X-ray state 
appears to increase by a factor of $\sim$9 relative to that in the lowest state. 
The relativistic broad Fe K$\alpha$ line is also stronger in the high state. 
Although the broad Fe K$\alpha$ line was significantly detected ($>3\sigma$) in only two \xmm observations 
(Obs-2 and Obs-3), their 99\% confidence contours are mutually exclusive, indicating a 
strong variability of the broad Fe K line component as well. 
 For the remaining two observations, the broad Fe K$\alpha$ intensity could be formally be zero at a 99$\%$ confidence level. 
 Taking into account the 99\% upper limit on the broad Fe K$\alpha$ intensity, the variability might be even larger. 
 Before the X-ray outburst, 
 NGC 1566 has also been observed by {\it Suzaku} on May 2012 
 when the X-ray continuum was in a historical low-flux state (Figure \ref{Fig:lcurve}).  
 While the {\it Suzaku} data gave sensitive measurements of the narrow Fe K$\alpha$ flux, 
 the broad Fe K$\alpha$ component was not significant detected at $<$90\% confidence \citep{Kawamuro2013}. 
 The red dot in Figure \ref{Fig:In_vs_Id_ct} represents narrow and broad Fe K$\alpha$ flux constrained by the {\it Suzaku} data, 
 which are consistent with our measurements within the statistical errors from Obs-1 which was in the similar low-flux state. 

From the baseline model, we could not place constraint on the inner radius($R_{in}$) for the broad Fe K$\alpha$ line emission. 
   Only the upper limit can be obtained as shown in Table \ref{Tab:xmm_result}. 
   With two \xmm observations in which the relativistic Fe K$\alpha$ line was detected, the inclination angles ($\theta_{obs}$) for the accretion disk can be tightly constrained 
   with a value of $41.37_{-1.91}^{+2.82}$ deg and $42.50_{-4.33}^{+10.88}$ deg, respectively. 
Next, we investigated the sensitivity of the broad Fe K$\alpha$ line flux in the XMM-Newton data to the Compton-reflection continuum.
The \textit{XMM-Newton} data did not constrain the reflection component well except for the first observation ($R=3.18_{-0.57}^{+0.60}$), and only the upper limit on the strength of the reflection component can be obtained for the other three observations when the X-ray continuum was higher (see Table \ref{Tab:xmm_result}). 
This would be expected since the \xmm data cover only a small portion of the Compton reflection continuum. 
In the next Section, we will present a joint fit to the \xmm and \nustar spectra. 
This is important in view of the extended sensitivity to higher energies provided by \nustar \citep[($\sim$3--80 keV,][]{2013ApJ...770..103H}, 
which allows us to put better constraint on the baseline model as well as the Compton reflection continuum. 
\begin{table*}[htbp!]
\centering
\fontsize{10}{30}
\caption{\label{Tab:rel_xill_result} Spectral Fitting Results for the quasi-simultaneous \textit{XMM-Newton} and \textit{NuSTAR} observations}
\begin{tabular}{cccc}
\hline
\hline
                 &Observation Date             & 2018-06-56                            & 2018-10-04\\\hline
Comp             & Parameter                   & Value                                 &    Value       \\\hline
                
\textit{relxill} & $q$                         & $3^f$                                 & $3^f$\\
                 & $\theta(degrees)$           & $42^f$                                & $42^f$\\
                 & $R_{in}(R_g)$               & $2.19 (<2.73)$                        & $14.47_{-11.54}^{+20.61}$\\
                 & $log(\xi)(erg\,cm\,s^{-1})$ & $<1.44$                               & $<2.03$\\
                 & $A_{Fe}$                    & $3.23_{-0.56}^{+0.68}$                & $2.84_{-0.70}^{+1.20}$\\
                 & $E_{cut}(keV)$              & $249.63_{-53.14}^{+143.52}$           & $224.54_{-74.95}^{+221.99}$\\
                 & $R_{ref}$                   & $0.36_{-0.045}^{+0.057}$              & $0.58_{-0.12}^{+0.06}$\\
                 & $\Gamma_{pn}$               & $1.84_{-0.028}^{+0.028}$              & $1.65_{-0.039}^{+0.044}$\\
                 & $\Gamma_{FPMA}$             & $1.75_{-0.031}^{+0.033}$              & $1.69_{-0.046}^{+0.044}$\\
                 & $\Gamma_{FPMB}$             & $1.75_{-0.015}^{+0.018}$              & $1.70_{-0.045}^{+0.026}$\\
                 & $norm$                      & $1.84_{-0.36}^{+0.38}\times\,10^{-4}$ & $1.87_{-1.00}^{+1.01}\times\,10^{-5}$\\
\textit{xillver} & $norm$                      & $1.59_{-0.39}^{+0.40}\times\,10^{-4}$ & $9.84_{-1.6}^{+2.0}\times\,10^{-5}$\\
\textit{const}   & $C_{FPMA}$                  & $1.16_{-0.030}^{+0.031}$              & $0.98_{-0.058}^{+0.056}$\\
                 & $C_{FPMB}$                  & $1.22_{-0.029}^{+0.032}$              & $0.99_{-0.060}^{+0.055}$\\\hline
                 & $L_{3.0-10.0keV}(10^{42} \,erg\, s^{-1})^a$               & 2.61                 & 0.776 \\
                 & $\chi^2$ /d.o.f             & 1195/1148                             & 774/772\\\hline
     
\end{tabular}
\begin{tablenotes}
\item[1]{\textbf{Notes}:Statistical errors and upper limits correspond to 90$\%$ confidence for one interesting parameter
($ \Delta\chi^2 $= 2.706). Frozen parameters are indicated by ``(f)''. $^a$ Intrinsic, rest-frame luminosities, corrected for all absorption components.}
\end{tablenotes}
\end{table*}

\subsection{Constraint on the Reflection Spectrum by Including the \nustar Data}\label{Sec:Var_ref}

\begin{figure*}[htbp!]
\includegraphics[scale=0.25]{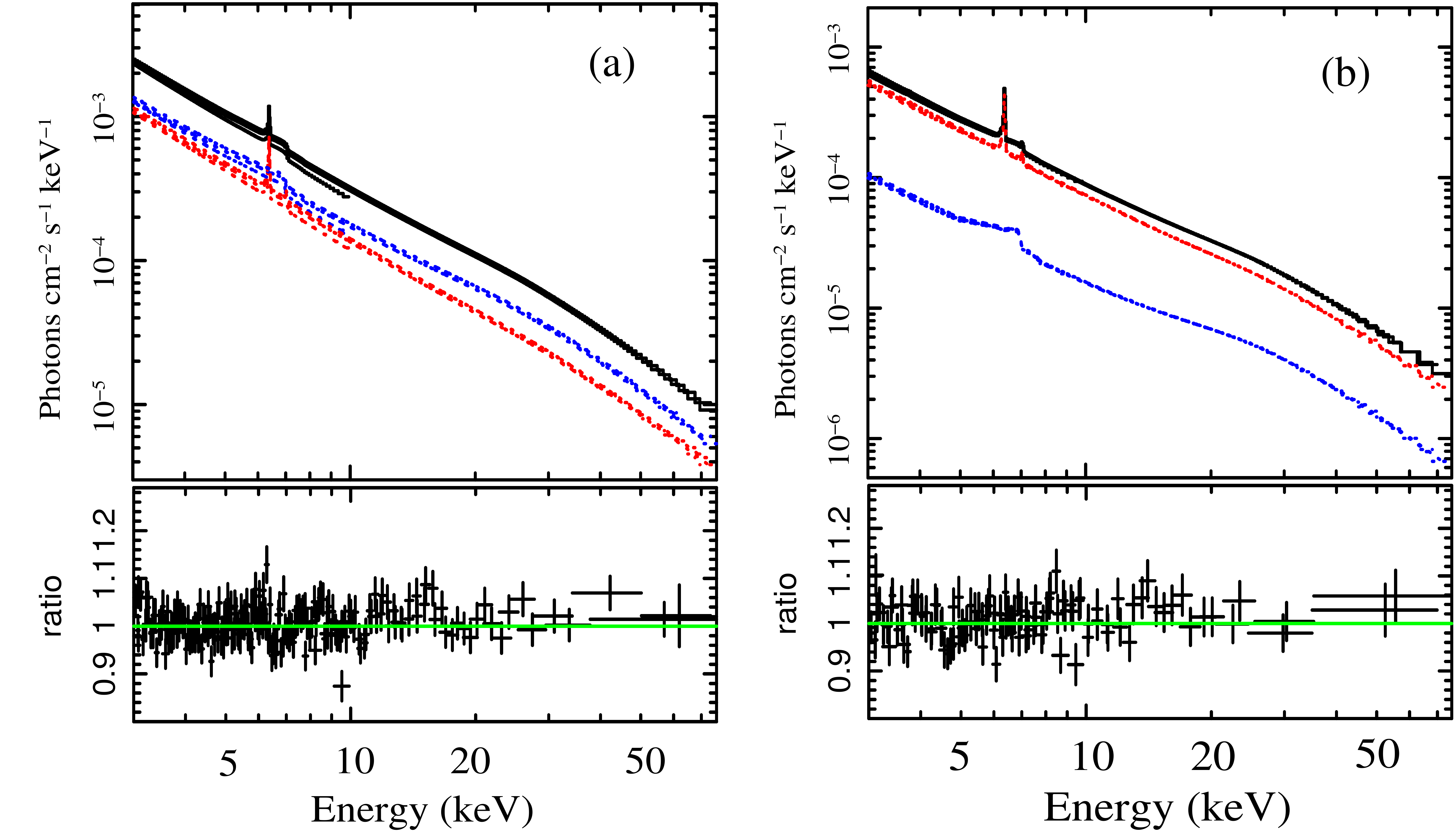}
\caption{The best-fitting reflection models to \textit{XMM-Newton} and NuSTAR data of
NGC 1566 which are parameterized by constant*wabs*(relxill+xillver). 
The relxill component is plotted in blue, while the dashed red line indicates the xillver component. 
The summed reflection model of relxill+xillver is shown in black. 
The corresponding data/model ratios are shown in the bottom panel. 
Panel (a) and (b) represents the fit to the \textit{XMM-Newton} and NuSTAR data observed on June 2018 and 
Oct 2018, respectively. \label{Fig:rel_xill_ufra}}
\end{figure*}     

Under the assumption that both the broad Fe K$\alpha$ line and the reflection continuum originate from reprocessing the same intrinsic 
X-ray continuum, a change in the Fe K$\alpha$ line flux should be accompanied by a corresponding change in the Compton reflection continuum. 
In this Section, we performed joint spectral fittings to the \textit{XMM-Newton} data (Obs-2 and Obs-3) and quasi-simultaneous \textit{NuSTAR} data. 
We have applied the {\tt relxill} model \citep{2013ApJ...768..146G} that is capable to calculate the disk reflection continuum and the associated 
relativistic Fe K emission lines self-consistently. To model the distant reflection emission and the associated narrow component of Fe K line, 
we have applied the {\tt xillver} model (without relativistic effect) \citep{2014MNRAS.444L.100D,2014ApJ...782...76G} independently. 
The parameters of {\tt xillver} were tied to those of {\tt relxill}, except for the flux normalization. 
We fixed the disk emissivity index, $q=3$, the value typical for a Shakura-Sunyaev disk \citep{1991ApJ...376...90L,2013MNRAS.430.1694D}. 
The black hole spin $a^*$ was fixed at the default value of 0.998 (the maximum spin parameter) as it cannot be well constrained
\footnote{When the black hole spin $a^*$ was allowed to vary in the fittings, it always approached to the value of 0.998.}.   
The outer disk radius $R_{out}$, was fixed at the default model value of 400 $r_g$ ($r_g=GM/c^2$ is the gravitational radius), 
and the inclination was fixed at the best-fit value derived from the phenomenological model, $\theta\sim 42^{\circ}$. 
The free model parameters are the power-law index, $\Gamma$, and the high energy cutoff, $E_{cut}$, the inner disk radius $R_{in}$, 
the ionization parameter, $\xi$, the iron abundance, $A_{Fe}$, and the reflection fraction, $R_{ref}$. 
The model, constant*wabs*(relxill+xillver), describes the data well, leaving no systematic data/model residuals 
with $\chi^2/dof$  = 1195/1148 and $\chi^2/dof$  = 774/772, for Obs-2 and Obs-3, respectively (see Figure \ref{Fig:rel_xill_ufra}). 
This suggests the fits are as good as the previous spectral fits with the baseline model (Section 3.2). 
The best fitting parameters are listed in the Table \ref{Tab:rel_xill_result}. 

 We obtained $\Gamma=1.84_{-0.028}^{+0.028}$ and $1.65_{-0.039}^{+0.044}$ for the power-law photon index for Obs-2 and Obs-3, 
 respectively, indicating a flattening of intrinsic continuum as the flux decreases. 
 The high energy cutoff ($ E_{cut}$) is consistent with each other, which was measured to be $249.63_{-53.14}^{+143.52}$ keV and $224.54_{-74.95}^{+221.99}$ keV 
 for the two observations. 
 For the ionization parameters of the accretion disk in the two observations, only upper limit can be obtained, with 
 log($\xi$)$<$1.44 and log($\xi$)$<$2.03, indicating a low to medium ionization disk. 
 The Fe element abundance($A_{Fe}$) in the accretion disk material and the inclination of the disk also play an important role 
 in describing the reflection spectrum of type I AGNs \citep{2006MNRAS.365.1067C,2014ApJ...782...76G}. 
 In our study, the inferred Fe abundance is consistent with each other between the two observations. 
 The reflection fraction was well constrained with $R=0.36_{-0.045}^{+0.057}$ and $0.58_{-0.12}^{+0.06}$, respectively, which is 
 consistent with the result of \citet{Kawamuro2013} ($R$=$0.45_{-0.10}^{+0.13}$). 
 This suggests that only a small amount of continuum flux illuminates accretion disk, explaining the low-to-medium ionization state of the disk. 
 These values are consistent with that obtained in other AGNs \citep{2019A&A...623A..12M,2020MNRAS.497.4213G}. 
We could only obtain loose constraints of $R_{\rm in}<2.73R_g$ for Obs-2 (June 2018), and $R_{\rm in}$=$14.47_{-11.54}^{+20.61}R_g$ for Obs-3 (Oct 2018). 
By constructing the contour plot between the inner radius ($R_{in}$) and the normalization of {\tt relxill} model, 
{we found that the 99\% confidence contours for the inner disk radius cannot be constrained. 
At only 68\% confidence, the $R_{\rm in}$ contours are mutually exclusive between the two observations, 
suggesting that the $R_{\rm in}$ variation is not significant.}

Finally, we also investigated whether the relativistic reflection spectrum (continuum plus broad Fe K$\alpha$ line) 
and distant reflection spectrum (continuum plus narrow Fe K$\alpha$ line) are both following the change in continuum flux between the 
two \xmm observations. We found that the flux of the relativistic reflection component has decreased from $1.84_{-0.36}^{+0.38}\times\,10^{-4}$ 
to $1.87_{-1.00}^{+1.01}\times\,10^{-5}$, while distant reflection component has decreased from $1.59_{-0.39}^{+0.40}\times\,10^{-4}$ to 
$9.84_{-1.6}^{+2.0}\times\,10^{-5}$. Considering the decrease in the continuum luminosity from $L_{3-10 keV}= 2.61\times 10^{42}\, erg\, s^{-1}$ to 
$L_{3-10 keV}= 7.76\times 10^{41}\, erg\, s^{-1}$ between the two \xmm observations, both data sets are indeed consistent with the 
reflection spectra that followed the intrinsic continuum variability. 

  \vspace{0.5cm}   
    \begin{figurehere}
 \centering
 \includegraphics[scale=0.2]{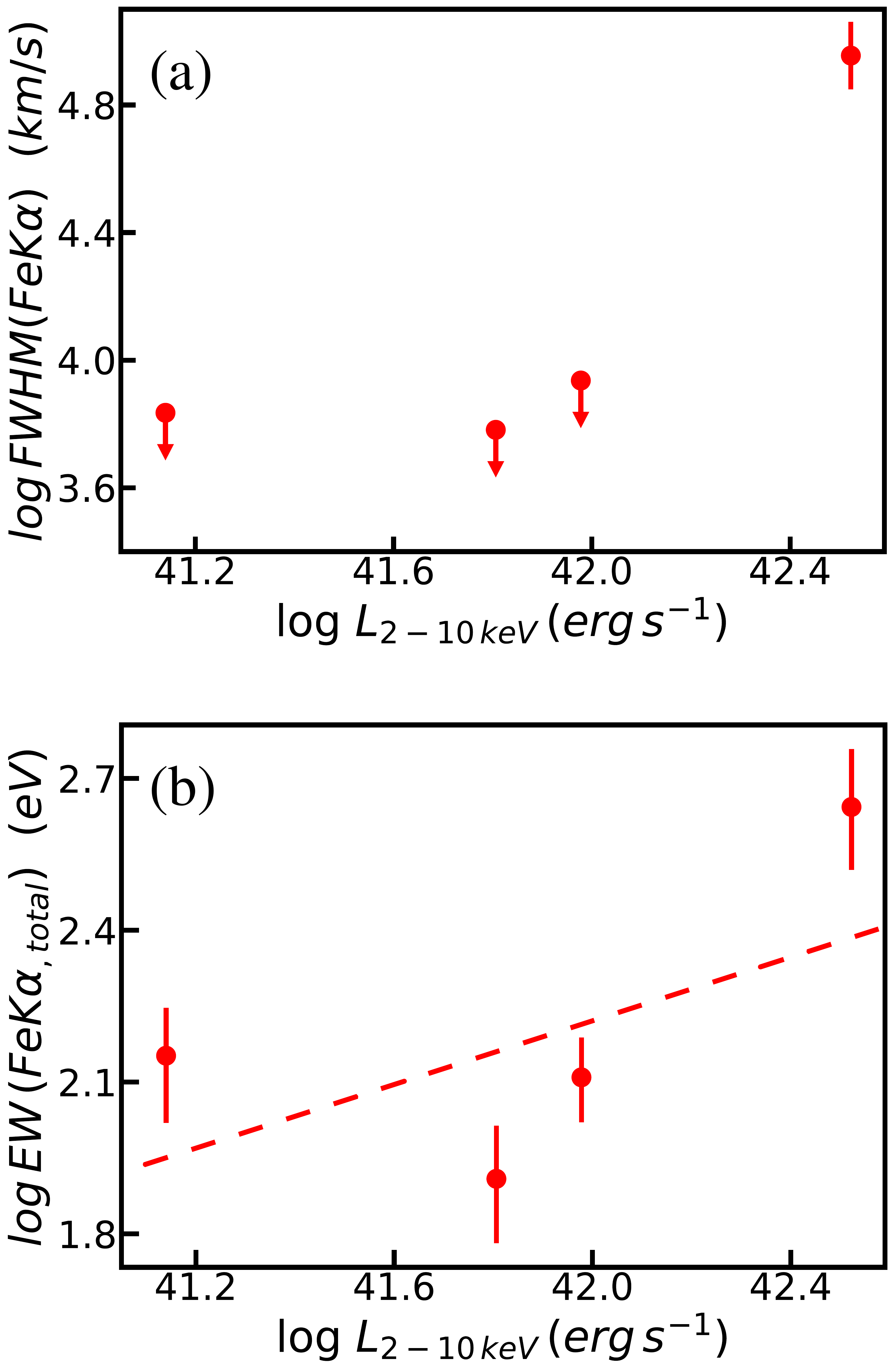}
 \caption{\label{Fig:broadline} 
 {\it Panel (a)}: The FWHM of Fe K$\alpha$ line fitted with a single Gaussian model versus the 2--10 keV luminosity for the four 
 \xmm observations. 
 {\it Panel (b)}: The same as (a) but for the EW of Fe K$\alpha$ line. 
 }
 \end{figurehere}

    \section{Discussion}

\subsection{Response of the Fe K$\alpha$ emission to the long-term X-ray continuum variability}

{As shown in Figure \ref{Fig:In_vs_Id_ct}, the intensity of broad and narrow Fe K$\alpha$ components in NGC 1566 
appears to change between different epochs. In this section, we will investigate in detail the relation of the Fe K line 
variability with the continuum. }
Figure \ref{Fig:broadline} (a) shows the relation between the FWHM of Fe K$\alpha$ line derived from the spectral fittings 
with a single Gaussian model and the 2--10 keV luminosity. It can be seen that at $L_{\rm 2-10 keV}<10^{42}$ erg s$^{-1}$ 
the FWHM of Fe K$\alpha$ line cannot be measured with an 90\% upper limit of $<$10000 km s$^{-1}$, 
while the FWHM reaches $\sim$0.3 c when the X-ray luminosity increased to $3.3\times10^{42}$ erg s$^{-1}$,  
indicating that the responsivity and/or structure of inner disk may have changed during the period of X-ray outburst. 
Figure \ref{Fig:broadline} (b) shows that the EW of relativistic Fe K$\alpha$ line component 
is also correlated with the X-ray luminosity. 
By fitting the \xmm and NuSTAR data with the reflection model {\tt relxill}, we found that the reflection 
fraction $R<1$ and is consistent with each other within errors between the two-epoch observations despite the 
3--10 keV flux varies by a factor of $>$3 (Section 3.4). 
This is in contrast to other AGNs where the reflection fraction has an inverse relation with the hard 
X-ray flux, which could be potentially explained by the light-bending effects \citep{Parker2014, Jiang2018}.  
In the lamp post geometry for the X-ray source, the lack of change in the reflection fraction in NGC 1566 suggests 
that the coronal height may have not changed during the period of X-ray outburst. 
The large amplitude X-ray variability is likely intrinsic, possibly caused by large change 
in the mass accretion rate, and the light-bending scenario appears not relevant for this AGN.
Note that besides NGC 1566, the only AGN that has been reported on the response of the broad Fe K$\alpha$ to continuum variations is NGC 2992 \citep{2010ApJ...713.1256S, 2020MNRAS.496.3412M}, 
which also displays the luminosity-dependent spectral CL phenomenon \citep{Guolo2021}. 

In Figure \ref{Fig:N_B_vs_C} (a), we plot the intensity of relativistic Fe K$\alpha$ line versus the 
2--10 keV continuum luminosity observed at different epochs. NGC 1566 is shown with red circles, while 
black circles represent NGC 2992 for comparison. 
We adopted the power-law function to analyze the emission line response to the change of continuum: 
$\rm log_{10}(f_{line})=\alpha log_{10}(L_{2-10 keV})+\beta$, where $\alpha$ is the power-law slope and $\beta$ 
is the intercept. The best-fitted slope is $\alpha=1.2\pm0.98$ and $\alpha=0.67\pm0.15$ for NGC 1566 and NGC 2992, respectively. 
Albeit with large errors, the variation rate of relativistic Fe K$\alpha$ line (the ratio of Fe K$\alpha$ variation to 
luminosity variation, characterized by the slope $\alpha$) seems different between the two AGNs. 
At the same X-ray luminosity, the strength of relativistic Fe K$\alpha$ line in NGC 1566 is larger than that in NGC 2992. 
This suggests the complex dependencies of the relativistic line emission on the geometry and physical properties of
the line-emitting material. 
The behaviour could be possibly explained by larger amount of gas or Fe abundance responsible for producing the Fe K$\alpha$ line in NGC 1566. 
Due to the lack of sensitive \nustar observations of NGC 2992 in the high state, we cannot constrain the Fe abundance  
from the X-ray spectral fittings, and hence test for the above latter scenario. 
 
Interestingly, we found the same trend of narrow Fe K$\alpha$ intensity in response to the continuum variation in NGC 1566, 
as shown in In Figure \ref{Fig:N_B_vs_C} (b). The best-fitted slope from the power-law function is $\alpha=0.57\pm0.13$. 
The narrow Fe K$\alpha$ line intensity can change by a factor of 1.5 for the corresponding continuum variation within a time interval as short as four months. 
This allows us to put the material responsible for producing the narrow Fe K$\alpha$ line at a distance scale to $<$0.1pc, 
comparable to the size of BLR which is $ \sim $ 0.01pc inferred for NGC 1566 \citep{1985ApJ...288..205A,2017MNRAS.470.3850D}. 
The rapidly variable narrow Fe K$\alpha$ line have also been reported in the AGN NGC 4151 
\citep[][]{Zoghbi2019}, which can be explained as originating from the inner BLR. 
 Although the variale narrow Fe K$\alpha$ line is detected in the AGN Mrk 841 \citep{2002A&A...388L...5P}, 
 the X-ray continuum remains constant during observations, hence it may have a different origin from NGC 1566. 
 On the other hand, we also found a relatively weaker response of the narrow Fe K$\alpha$ line to the continuum variation in NGC 2992 
 ($\alpha=0.30\pm0.07$). 
 \citet{Ghosh2021} has suggested that the narrow Fe K$\alpha$ line could originate from the reprocessor at a distance of 0.3--1.0 pc, 
 consistent with that of torus reflection. 

\begin{figurehere}
\centering
\includegraphics[scale=0.22]{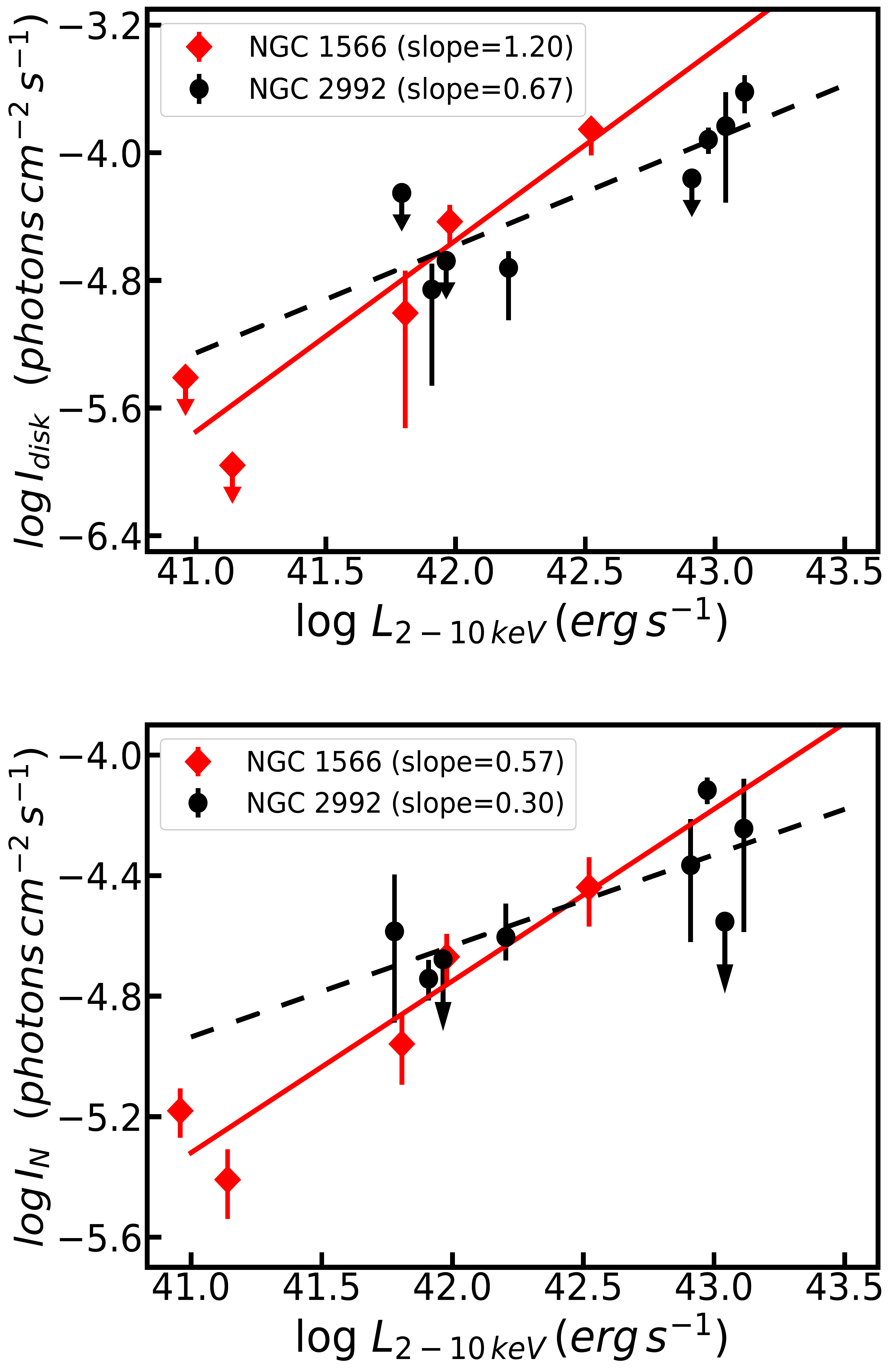}
\caption{ \label{Fig:N_B_vs_C}\textit{Upper panel}: The broad Fe K$\alpha$ intensity ($I_{disk}$) versus the primary X-ray continuum 
luminosity in the 2-10 keV. \textit{Lower panel}: The narrow Fe K$\alpha$ intensity ($I_{N}$) versus the primary X-ray continuum luminosity. The data for NGC 2992 comes from \citep{2007PASJ...59S.283Y, 2010ApJ...713.1256S} and our spectral fittings to the new X-ray data taken from \xmm and \nustar observations since 2010. 
The fitting results with power-law function to the data are displayed by the dashed lines.}
\end{figurehere}
 
 It is well known that the EW of the narrow Fe K$\alpha$ line is 
 anti-correlated with X-ray continuum luminosity for a sample of AGNs, so-called X-ray Baldwin Effect 
 \citep[e.g.,][]{Jiang2006, Bianchi2007, Shu2011}. 
 In Figure \ref{Fig:xraybe}, we plot the X-ray observations of NGC 1566 and NGC 2992, 
 and find a clear decrease in the narrow Fe K$\alpha$ EW as the luminosity increases, with 
 $EW\propto L^{-0.35\pm0.08}$ and $EW\propto L^{-0.66\pm0.09}$, respectively. 
 Therefore, the anti-correlation between EW and continuum luminosity is much stronger in NGC 2992. 
 This result could be explained by the fact that the narrow Fe-K line emission in NGC 2992 originates from a 
 region farther from the nucleus, so has a weaker response to the continuum variation within several years. 
 By comparing with the AGN sample with the {\it Chandra} high-energy grating observations \citep{Shu2010b}, 
 both NGC 1566 and NGC 2992 have steeper slope for the anti-correlation between EW and continuum luminosity, 
 suggesting that the observed X-ray Baldwin Effect could partially be attributed to the intrinsic continuum 
 variation in AGNs \citep[e.g.,][]{Jiang2006, Shu2012}. 
 It should be noted that most previous studies on the X-ray Baldwin Effect 
 did not perform the detailed deconvolution of the narrow and broad Fe K$\alpha$ line component, 
 so the results could be affected by the variability of underlying broad Fe K$\alpha$ line if presents. 
 X-ray monitoring observations of intensely varying AGNs, especially those CL AGNs, on timescales of days to years,  
 are crucial to study the response of both the relativistic and narrow Fe K$\alpha$ line emission to the changes in the X-ray continuum, 
 which can provide new insight into the properties and origins of all of the Fe K$\alpha$ emission line, as well as 
 the circumnuclear structures for producing them.

 \vspace{0.01cm}

 \begin{figurehere}
\centering
\includegraphics[scale=0.26]{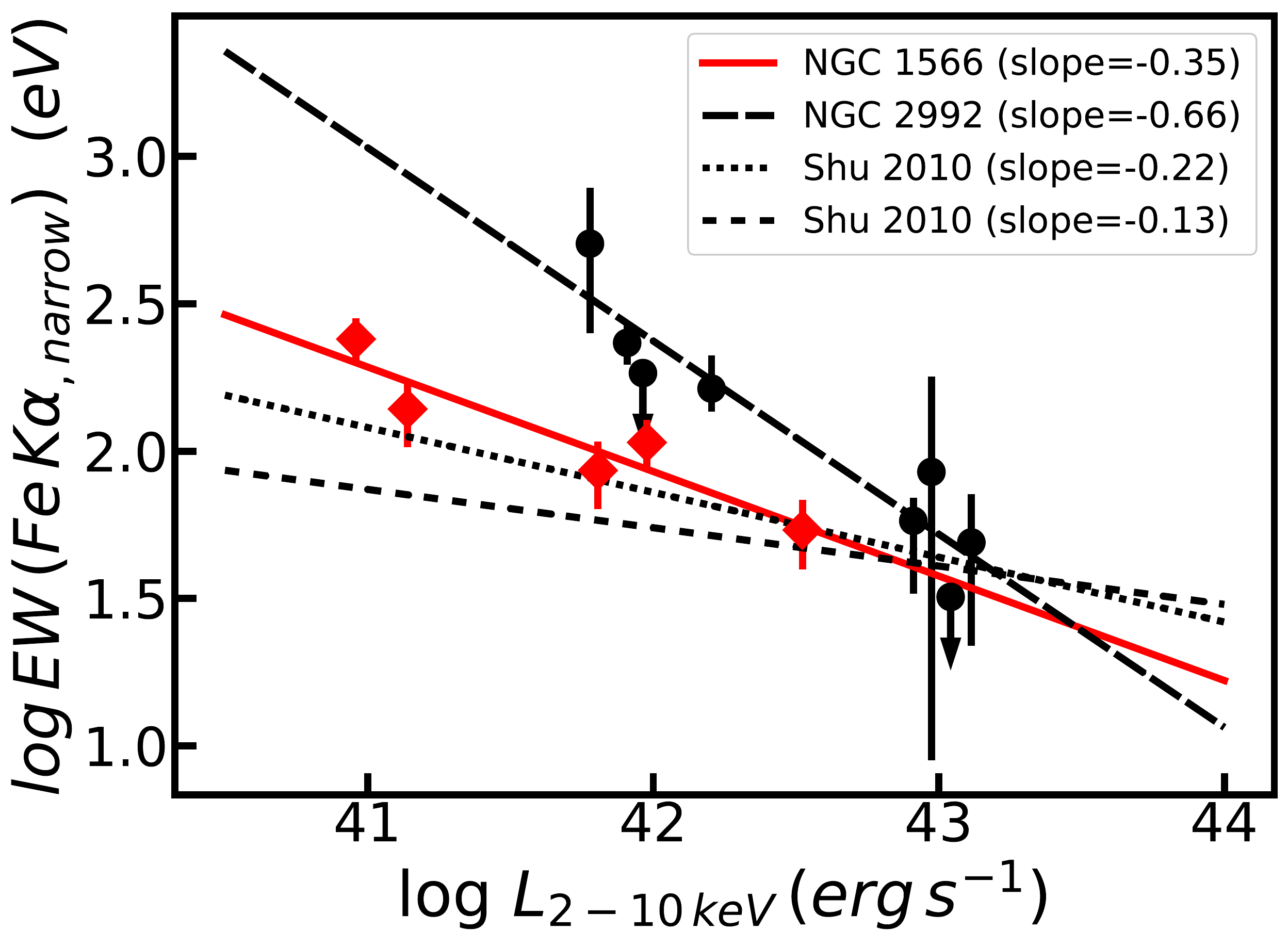}
\caption{\label{Fig:xraybe} 
The inverse correlation between the narrow Fe K$\alpha$ line EW and the 2--10 keV luminosity 
for NGC 1566 (red filled circles) and NGC 2992 {(black filled circles).} 
For comparison, we also present the best-fit slopes for the AGN sample observed by {\it Chandra} high-energy gratings \citep{Shu2010b}. 
The black dotted line was constructed from the fit to the sample made from individual {\it Chandra} observations (``per observation''), 
while the dashed line was for the sample made from spectra combining multiple observations for a given source (``per source''). 
}
\end{figurehere}

\subsection{Implications on the AGN state transition by the X-ray spectral variability}

 As shown in Figure \ref{Fig:xmm_p_m_ra}, there is a variable soft X-ray excess emission (below $\sim2$ keV) between the four \xmm observations. 
 If modelled by a power-law component, we obtained a photon index of $\Gamma=3.21_{-0.12}^{+0.42}$ in the highest state observed on June 2018. 
 The strength of soft X-ray excess appears to be correlated with the X-ray continuum luminosity. 
The origin of soft X-ray excess is still a controversial topic. 
One popular possibility is that there are two Comptonizing regions. 
The Comptonization by an optically thick warm corona could account for, at least partially, 
the observed soft X-ray excess emission in AGNs \citep[e.g.,][]{Done2012, Shu2017, Jin2021}, 
and the Comptonization by high-temperature optically thin corona describes the standard X-ray power-law
emission above 2 keV. 
In this model, the soft X-ray excess is related to the disk emission, so its strength can be 
powered by mass accretion. 
We calculated the Eddington ratio ($\lambda_{Edd}$=$L_{bol}/L_{Edd}$) for the four \xmm observations using the luminosity 
in the 2-10 keV band and a black hole mass of $ \sim 10^7 $ \citep{2002ApJ...579..530W}. A bolometric correction factor of 20 \citep{2009MNRAS.399.1553V} 
was assumed to extrapolate the X-ray luminosity to the bolometric luminosity. 
For Obs-1 and Obs-2, we obtained the same Eddington ratios as reported in \citet{2019MNRAS.483L..88P}. 
To investigate the $\lambda_{Edd}$-dependence of the soft-excess emission, we quantified the strength 
of the soft excess as the ratio of flux at soft X-ray band ($L_{\rm S, total}$) to the powerlaw extrapolation 
of hard X-ray component to the same band ($L_{\rm S, powerlaw}$). 
Since the soft excess emission is variable in flux, we parametrized the strength of the soft excess 
at three soft X-ray bands, namely, 0.5--1.0 keV, 0.5--1.5 keV and 0.5--2.0 keV, respectively.

{Figure \ref{Fig:L/Ledd} (upper panel)} displays the correlation of soft excess ($L_{\rm S, total}$/$L_{\rm S, powerlaw}$) with the Eddington ratios. 
It can be seen that the strength of soft X-ray excess is strongly dependent of Eddington ratio. 
At $L_{bol}/L_{Edd}<0.01$, there is no or little soft excess emission at energies below $\sim$2 keV. 
Conversely, soft excess emission is observed at all epochs when the accretion rate is across a critical value of $L_{bol}/L_{Edd}\sim0.01$. 
This could be explained as due to a state change associated with the spectroscopic changing-look phenomenon \citep{2019MNRAS.483..558O, 2020MNRAS.498..718O}, 
similar to that observed in CL AGN Mrk 1018 \citep{2019MNRAS.483L..88P} and Mark 590 \citep{Denney2014}. 
Our results are broadly consistent with the recent study by \citet{Jana2021}, confirming the variable soft excess emission 
and its dependence on the Eddington ratios. 
The observations of NGC 1566 seem to support the scenario proposed by \citet{2018MNRAS.480.3898N} that AGNs will likely 
experience changing-look phenomenon when their luminosity variations cross the state transition boundary 
of $L_{bol}/L_{Edd}$ at a few percent. 

\begin{figurehere}
\centering
\includegraphics[scale=0.23]{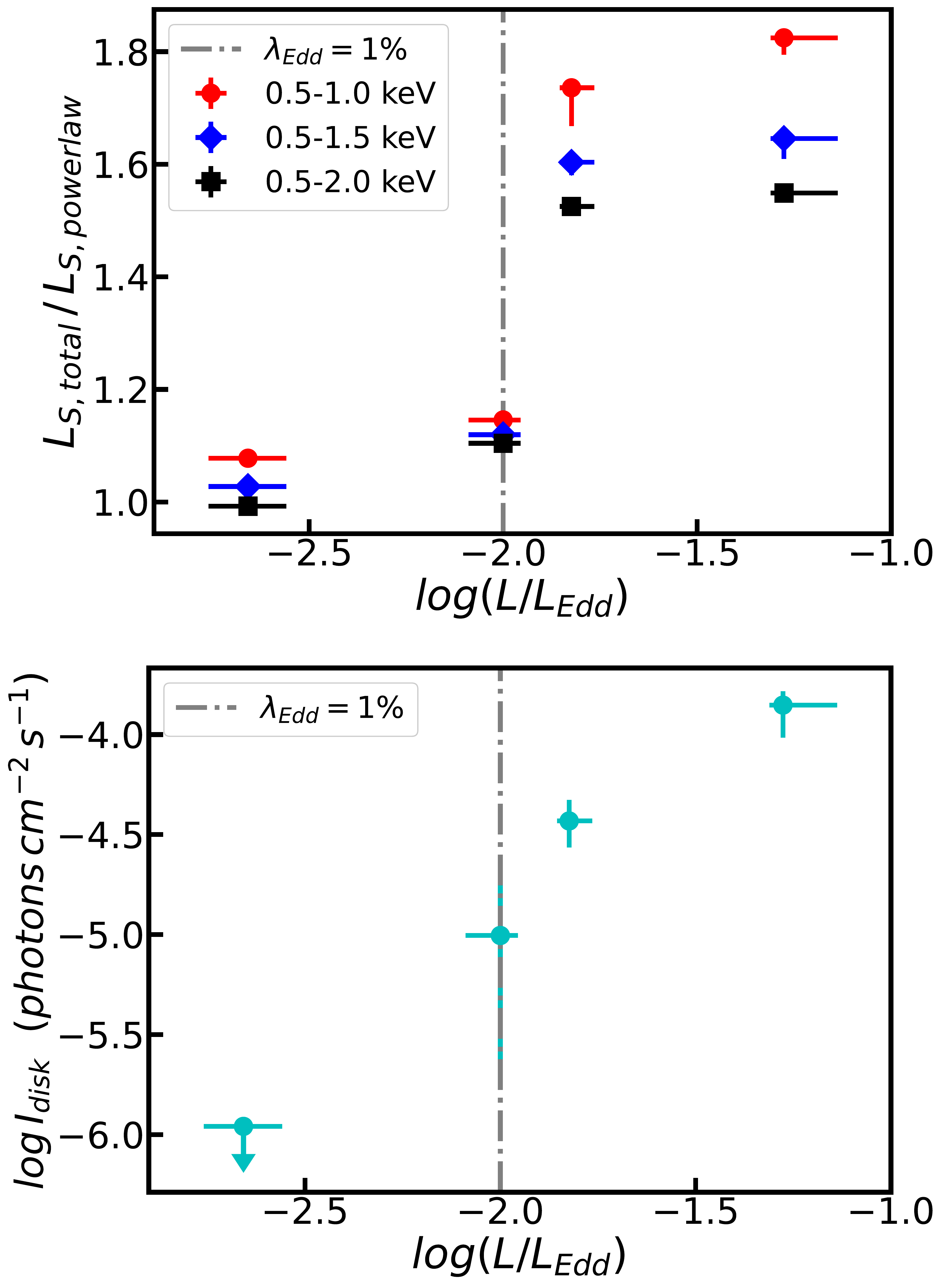}
\caption{\label{Fig:L/Ledd} 
{The dependence of the soft-excess strength (upper panel) and broad Fe K$\alpha$ line intensity (lower panel) on the Eddington ratio ($L_{bol}/L_{Edd}$). }
The strength of the soft-excess is parametrized by the ratio of total luminosity in the 
soft X-ray band to that extrapolated from the hard X-ray powerlaw component. 
We adopted soft X-ray band of 0.5--1.0 keV, 0.5--1.5 keV and 0.5--2.0 keV to 
describe the corresponding soft-excess emission, respectively. 
Errors here refer to 90$\%$ confidence intervals.}
\end{figurehere}

{Figure \ref{Fig:L/Ledd} (lower panel) shows that the intensity of the broad Fe K$\alpha$ line 
is also dependent of Eddington ratio, if the bolometric luminosity can be traced by the X-ray luminosity. 
In the model of \citet{2018MNRAS.480.3898N}, if AGNs cross a few per cent in their $L/L_{Edd}$, they 
would show state transition due to disk evaporation/condensation associated with the decrease/increase 
in accretion rate. This appears similar to the soft-to-hard state transition in black hole binaries at $L/L_{Edd}\sim0.02$, 
 below which the inner disk could evaporate into an advection dominated accretion flow \citep{Done2007}. 
 Indeed, evidence of disk truncation at low accretion states has been found with recent XMM and NuSTAR observations of black hole binaries 
 \citep{Xu2020a, Xu2020b}. 
In this context, an increase in the flux of broad Fe K line emission during the outburst of NGC 1566 could be 
explained by the condensation of the inner accretion disk and its intensified illumination by an X-ray corona. 
Unfortunately, we could only obtain loose constraints on the radius of inner disk ($R_{\rm in}$) from spectral fittings 
(Section 3.4), 
hence only marginal evidence for the decrease in the inner disk during the outburst state could be inferred. 

} 

Finally, it should be noted that though variable soft X-ray excess has been observed in many CL AGNs, the observation of variable broad, relativistic 
Fe K$\alpha$ line emission is rare. NGC 1566 and NGC 2992 are the two CL AGNs known so far in which broad Fe K$\alpha$ line emission in response to 
large amplitude X-ray continuum variability is observed. The obscuration seems not be the reason for the non-detections of broad Fe K$\alpha$ line in 
other CL AGNs, as NGC 2992 has shown variable absorption along the light of sight with $N_H=1-3\times10^{22}$ cm$^{-2}$ \citep{2010ApJ...713.1256S}. 
It is possible that EW of broad Fe K$\alpha$ line is not large enough so that its contrast against the continuum is not high 
enough to constrain the line profiles \citep{Giustini2017}. In fact, it still remains unclear why only a faction of AGNs have shown broad Fe K$\alpha$ line 
in their X-ray spectra, even with sufficient signal-to-noise ratio \citep[e.g.,][]{Nandra2007, Mantovani2016}. 
As shown in Appendix, we performed the time-lag analysis for NGC 1566, based on the lightcurve in Obs-2 which 
shows the source is variable on a time-scale as short as a few kilosecond. 
Although we found a hard lag at low-frequency which can be understood with the propagating fluctuations model, 
{no reverberation signal is observed at higher frequency. }
Therefore, the Fe K reverberation commonly observed in AGNs may probe the disk reflection at 
much different time-scales from the long-term variability of relativistically broadened Fe K$\alpha$ line emission 
in response to the continuum variations in NGC 1566.  
We propose that 
NGC 1566 and similar changing-look AGNs with both variable soft X-ray excess and broad Fe K$\alpha$ line emission 
provide a unique laboratory to probe the central engine of AGN, and future X-ray monitoring observations with higher cadence 
are encouraged. 

\section{Acknowledgments} 
This research has made use of data obtained from {\it XMM-Newton}, an ESA
science mission with instruments and contributions directly funded by ESA and NASA, 
and from the NuSTAR, a mission managed by the Jet Propulsion Laboratory, and funded by NASA.
	This research has made use of the HEASARC online data archive services, supported by NASA/GSFC.
        This work is supported by Chinese NSF through grant Nos. 11822301, 12192220, 12192221, 12033006, and 11833007. 
\end{multicols}

\begin{multicols}{2}

\end{multicols}

	\newpage
	\clearpage

\begin{multicols}{2}

	\section{Appendix}
	\subsection{X-ray time lag analysis}
	
\begin{figure*}[htbp!]
\centering
\includegraphics[scale=0.3]{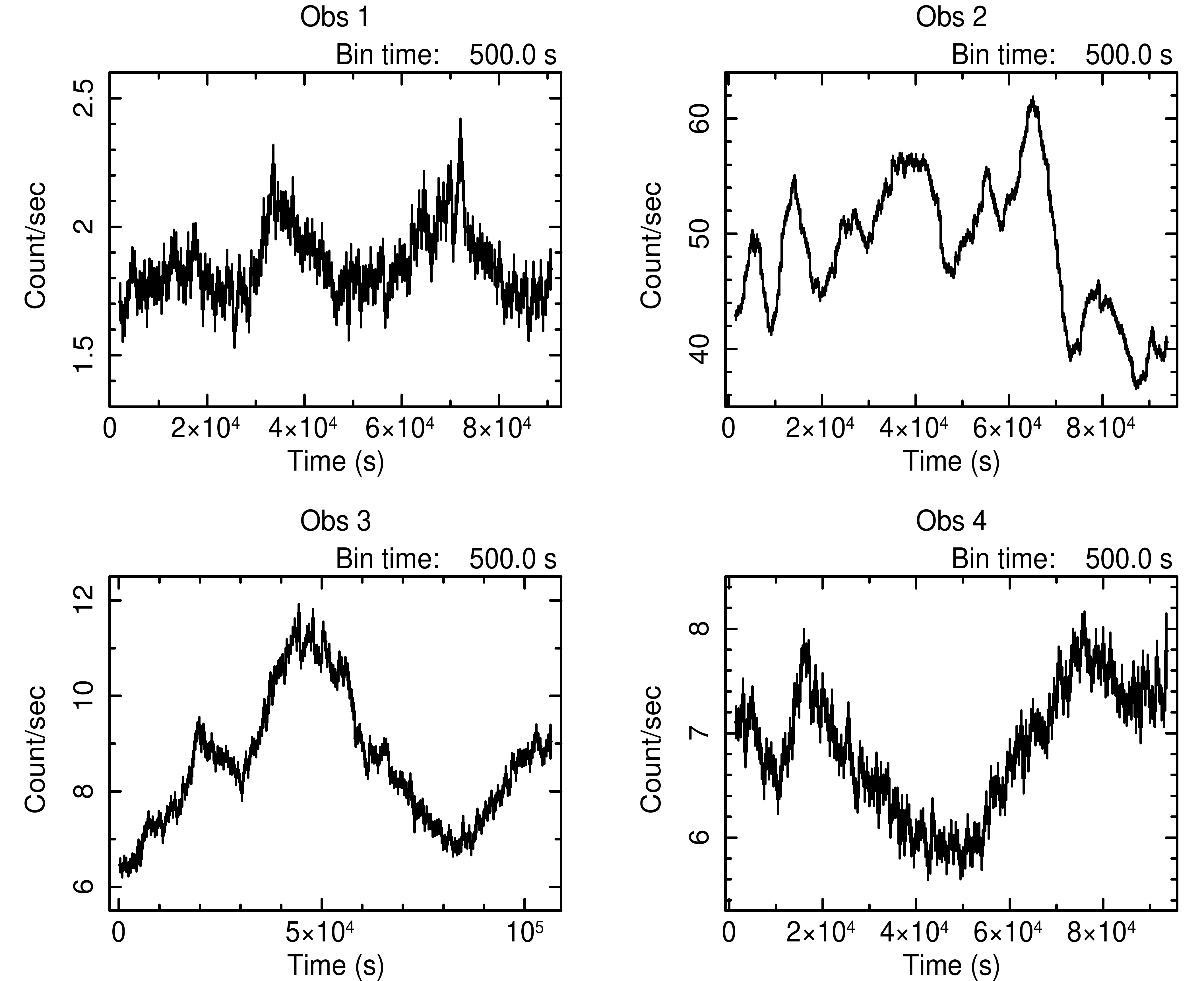}
\caption{\label{Fig:lc} 
The 0.3-10 keV light curve with 500s bins for the four \xmm observations of NGC 1566. 
}
\end{figure*}	
	
{Figure \ref{Fig:lc} shows the 0.3--10 keV light curve for the data taken from the four \xmm
observations. All the light curves were background-subtracted, using the same region as that to extract spectra, 
and then corrected for instrumental effects using the SAS tool {\tt epiclccorr}.
The data are grouped into bins with interval widths of 500s. The light curve shows variability 
by a factor of $\simlt$2 with a time-scale of tens of kiloseconds. It can be seen that the source is relatively 
more variable during the period of outburst (XMM Obs-2) on a shorter time-scale of kiloseconds. Therefore, we used the 
lightcurve from Obs-2 to perform the Fourier time lag analysis, following the standard procedures \citep{Uttley2014,Nowak1999}. 
We first constructed the lag–frequency spectrum between 0.3–1 keV and 1-4 keV to search for characteristic frequency at 
which the time lag might be present. These two energy bands are chosen because the X-ray emission between 0.3-1 keV usually characterizes soft X-ray excess, 
while that at 1-4 keV is dominated by the harder power-law component. The result is shown in the left panel of Figure \ref{Fig:lag}. 
We find a clear evidence of positive lags, i.e., the hard band emission lags behind the soft band emission below the frequency $\sim3\times10^{-4}$Hz. 
No negative lags at higher frequency were found which can trace the variability information on the inner region of accretion flows. 
In order to get a lag-energy spectrum, the light curves were constructed in ten energy bins between 0.3 and 10 keV with 1 keV bin width 
(0.3-1keV is used as one energy bin). The time lag was then computed for the light curve in each energy bin of interest 
relative to the reference light curve between 0.3–10 keV. The 0.3–10 keV light curve was used as a reference because it helps to maximize the signal to noise. 
In order to eliminate the correlated errors, we removed the band of interest from the reference band when computing the time lags. 
The resulting lag-energy spectrum is shown in right of panel of Figure \ref{Fig:lag}.  
It can be seen that there is a hard lag that increases with increasing energy. 
Such a hard lag at low-frequency is commonly seen in Seyfert galaxies \citep{2016MNRAS.462..511K}, 
which can be understood within the framework of the propagating fluctuations model. 
Therefore, our time lag analysis suggests no X-ray reverberation signal arising from gravitationally 
redshifted Fe K$\alpha$ photons reflected off the inner accretion flow for NGC 1566. 

\begin{figure*}[ht]
\centering
\includegraphics[width=\linewidth]{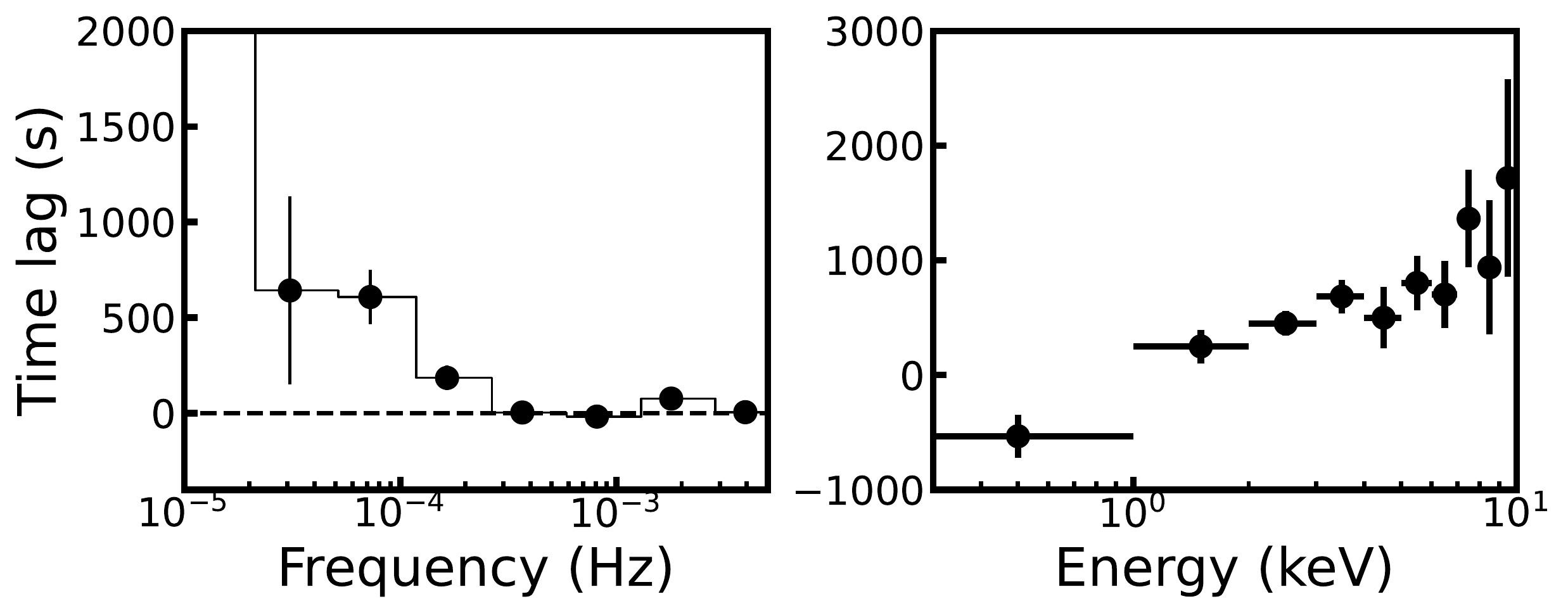}
\caption{\label{Fig:lag} 
Left: the lag–frequency spectrum between 0.3–1 keV and 1–4 keV for the XMM Obs-2 data. Positive lags at \textless  $3\times10^{-4}$Hz indicate hard lags. 
Right: the low-frequency lag-energy spectrum (see the text for details).  
Error bars are at the 1$\sigma$ level.}
\end{figure*}

} 	 
\end{multicols}

\end{document}